\documentclass[manuscript,screen]{acmart}

\AtBeginDocument{%
  \providecommand\BibTeX{{%
    \normalfont B\kern-0.5em{\scshape i\kern-0.25em b}\kern-0.8em\TeX}}}

\setcopyright{acmcopyright}
\copyrightyear{2018}
\acmYear{2018}
\acmDOI{10.1145/1122445.1122456}

\usepackage{bm, bbm}
\usepackage{amsfonts}
\usepackage{mathtools}
\usepackage{amsmath}
\usepackage{mathtools}
\usepackage{array}
\usepackage{bold-extra}

\usepackage{mathtools}

\usepackage{graphicx}
\usepackage{float}
\floatstyle{plaintop}
\usepackage{cleveref}

\usepackage{braket}

\usepackage{tikz}
\usetikzlibrary{shapes,positioning}
\tikzset{ myell/.style={ draw, rounded rectangle } }

\usepackage{amsthm}
\newtheorem{my_definition}{Definition}

\newtheorem{my_lemma}{Lemma}
\newtheorem{my_theorem}{Theorem}

\newtheorem{appendix_lemma}{Lemma A}

\newtheorem{appendix_theorem}{Theorem A}

\usepackage{tabularx}
\usepackage{ltablex}

\usepackage{footnote}

\usepackage{braket}

\usepackage{tikz}

\usepackage{algorithm}
\usepackage[noend]{algpseudocode}
\makeatletter
\def\algbackskip{\hskip-\ALG@thistlm}
\makeatother
\usepackage{algpseudocode}

\begin{document}

\title{Quantum Machine Learning Algorithm for Knowledge Graphs}


\author{Yunpu Ma}
\email{cognitive.yunpu@gmail.com}
\affiliation{
	\institution{Ludwig Maximilian University of Munich}
}

\author{Volker Tresp}
\email{volker.tresp@siemens.com}
\affiliation{
	\institution{Ludwig Maximilian University of Munich $\&$ Siemens CT}
}

\begin{abstract}
Semantic knowledge graphs are large-scale triple-oriented databases for knowledge representation and reasoning. Implicit knowledge can be inferred by modeling and reconstructing the tensor representations generated from knowledge graphs. However, as the sizes of knowledge graphs continue to grow, classical modeling becomes increasingly computational resource intensive. This paper investigates how to capitalize on quantum resources to accelerate the modeling of knowledge graphs. In particular, we propose the first quantum machine learning algorithm for making inference on tensorized data, e.g., on knowledge graphs. Since most tensor problems are NP-hard \cite{hillar2013most}, it is challenging to devise quantum algorithms to support that task. We simplify the problem by making a plausible assumption that the tensor representation of a knowledge graph can be approximated by its low-rank tensor singular value decomposition, which is verified by our experiments. The proposed sampling-based quantum algorithm achieves speedup with a polylogarithmic runtime in the dimension of knowledge graph tensor.
\end{abstract}



\maketitle

\section{Introduction}

Semantic knowledge graphs (KGs) are graph-structured databases consisting of semantic triples \emph{(subject, predicate, object)}, where subject and object are nodes in the graph, and the predicate is the label of a directed link between subject and object. An existing triple normally represents a fact, e.g., \textit{( California, located\_in, USA)} and missing triples stand for triples known to be false (closed-world assumption) or with an unknown truth value. In recent years a number of sizable knowledge graphs have been built, such as \textsc{Freebase}~\cite{bollacker2008freebase}, \textsc{Yago}~\cite{suchanek2007yago}, etc. The largest knowledge graph, e.g., Google's Knowledge Vault~\cite{dong2014knowledge}, contains more than $100$ billion facts and hundreds of millions of distinguishable entities.

An adjacency tensor can represent a knowledge graph with three dimensions: One stands for subjects, one for predicates, and one for objects. More precisely, we let $ \chi \in \{ 0, 1 \}^{d_1 \times d_2 \times d_3} $ denote the semantic tensor of a knowledge graph, where $d_1$, $d_2$, and $d_3$ represent the number of subjects, predicates, and objects, respectively. An entry $x_{s p o}$ in $\chi$ assumes the value $1$ if the semantic triple $(s, p, o)$ is known to be true, while it assumes the value $0$ if the triple is false or missing. Machine learning aims to infer the truth value of triples, given the knowledge graph triples were known to be true. Popular learning-based algorithms for modeling KGs are based on a factorization of the adjacency tensor such as the Tucker tensor decomposition, PARAFAC, RESCAL~\cite{nickel2011three}, or compositional models such as DistMult~\cite{yang2014embedding} and HolE~\cite{nickel2016holographic}.

The vast number of facts and entities makes it particularly challenging to scale learning and inference algorithms to perform inference on the entire knowledge graph. This paper aims to use quantum computation to design algorithms that can dramatically accelerate the inference task. Thanks to the rapid development of quantum computing technologies, quantum machine learning~\cite{biamonte2017quantum} is becoming an active research area that attracts researchers from different communities. In general, quantum machine learning exhibits great potential for accelerating classical algorithms, e.g., solving linear systems of equations~\cite{harrow2009quantum}, supervised and unsupervised learning~\cite{wiebe2014quantum}, support vector machines~\cite{rebentrost2014quantum}, Gaussian processes~\cite{das2018continuous}, non-negative matrix factorization~\cite{du2018quantum}, recommendation systems~\cite{kerenidis2016quantum}, etc.

Most of the aforementioned quantum machine learning algorithms contain subroutines for singular value decomposition, singular value estimation, and singular value projection of data matrices prepared and presented as quantum density matrices.  We show that the tensor factorization algorithm presented in this paper, which uses existing quantum algorithms as subroutines, has a polylogarithmic runtime complexity. However, unlike matrices, most tensor problems are NP-hard, and there is no current quantum algorithm that can handle tensorized data. Therefore, to understand the difficulties of designing quantum machine learning algorithms on tensorized data, e.g., data derived from a vast relational database, we need first to answer the following questions:

(1) Under what conditions can we infer implicit knowledge from an incomplete knowledge graph by reconstructing it via classical algorithms; (2) Does there exist an analogous tensor singular value decomposition method that we can map to a quantum algorithm? (3) Assuming that the knowledge graph has global and well-defined relational patterns, can the tensor SVD of a subsampled semantic tensor approximate the original tensor. Mainly, after projecting onto the lower-rank space, previously unobserved truth values of semantic triples might be boosted? (4) If all the above conditions are fulfilled, how can we design a quantum algorithm which projects the tensorized data onto lower-rank space to reconstruct the original tensor?

The first part of this paper contributes to the classical theory of binary tensor sparsification. As a novel contribution, we derive the first binary tensor sparsification condition under which the original tensor can be approximated by the truncated or projected tensor SVD of its subsampled tensor. The second part focuses on developing the quantum machine learning algorithm. To handle the tensorized data, we first explain a quantum tensor contraction subroutine. We then design a quantum learning algorithm on knowledge graphs using quantum principal component analysis, quantum phase estimation, and quantum singular value projection. We study the runtime complexity and show that this sampling-based quantum algorithm provides acceleration w.r.t. the size of the knowledge graph during inference.

\subsection{Related Work}

In this section, we discuss recent work on quantum machine learning for big data. It is commonly believed that the quantum recommendation system (QRS) proposed in \cite{kerenidis2016quantum} will potentially be one of the first commercial applications of quantum machine learning. The quantum recommendation system provides personalized recommendations to individual users according to a preference matrix $ A $ with runtime $ \mathcal{O} ( \mathrm{poly}(k) \mathrm{ polylog } ( mn ) ) $, where $ m \times n $ is the size of the preference matrix $ A $ which is assumed to have a low rank-$k$ approximation. On the other hand, a recent breakthrough made by \cite{tang2019quantum} shows that by dequantizing the quantum recommendation algorithm, a classical machine learning algorithm can achieve the same acceleration if the classical algorithm has access to a data structure which resembles the one required in the QRS. However, as commented by the authors of \cite{kerenidis2016quantum} in \cite{kerenidis2018q}, this new classical algorithm is based on the FKV methods \cite{frieze2004fast} has a much worse polynomial dependence on the rank of the preference matrix and a dramatic slowdown dependence on a predefined precision parameter, making it completely impractical. Therefore, it remains an open question to find the corresponding dequantized classical algorithms for machine learning on tensorized data that are polylogarithmic in dimension as the proposed quantum algorithm.

A recent work \cite{gu2019quantum} that presents a quantum algorithm for higher-order tensor singular value decomposition (HOSVD) \cite{de2000multilinear}. The quantum HOSVD algorithm decomposes a $m$-way $n$-dimensional tensor into a core tensor and unitary matrices with computational complexity $ \mathcal{O} ( m n^{3/2} \log^{m} n ) $. It provides an acceleration compared with the classical HOSVD with complexity $ \mathcal{O} ( m n^{m+1} ) $. Note that the polynomial dependence of the complexity on the tensor dimension comes from the quantum subroutines since the quantum HOSVD reconstructs the core tensor and unitary matrices explicitly. In contrast, our quantum tensor SVD method doesn't estimate singular values and unitary matrices explicitly. Instead, it samples results from a projected tensor under the assumption that the tensorized data has a low-rank orthogonal approximation. Hence, it provides a polylogarithmic dependence on the tensor dimension. Almost all classical knowledge graph embedding and completion methods are based on HOSVD or similar techniques. Take Tucker \cite{tucker1966some} as an example (since many other methods, such as DistMult, Rescal, CeomplEx, are special cases of Tucker); it is based on the Tucker decomposition of the binary tensor representation of knowledge graph triples (i.e., $m=3$). Hence, we expect our quantum tensor SVD method to be much faster than all the classical approaches. 

\section{Tensor Singular Value Decomposition}

First, we recap the singular value decomposition (SVD) of matrices. Then we introduce tensor SVD and show that a given tensor can be reconstructed with a small error from the low-rank tensor SVD of the subsampled tensor. Other tensor decomposition algorithms, e.g., higher-order tensor SVD~\cite{de2000multilinear}, will not be considered in this work since designing their quantum counterparts can be much more involved.

\textbf{SVD $\quad$ } Let $A \in \mathbb{R}^{m \times n}$, the SVD is a factorization of $A$ is the form $A = U \Sigma V^{ \intercal }$, where $ \Sigma $ is a rectangle diagonal matrix singular values on the diagonal, $U \in \mathbb{R}^{m \times m}$ and $V \in \mathbb{R}^{n \times n}$ are orthogonal matrices with $U^{ \intercal } U = U U^{ \intercal } = I_m $ and $ V^{ \intercal } V = V V^{ \intercal } = I_n$.

\textbf{Notations for Tensors $\quad$} A $N$-way tensor is defined as $\mathcal{A} = ( \mathcal{A}_{i_1 i_2 \cdots i_N}) \in \mathbb{R}^{d_1 \times d_2 \times \cdots \times d_N}$, where $d_k$ is the $k$-th dimension. Given two tensors $\mathcal{A}$ and $\mathcal{B}$ with the same dimensions, the inner product is defined as $ \langle \mathcal{A}, \mathcal{B} \rangle_F :=   \sum_{i_1 = 1}^{d_1} \cdots \sum_{i_N = 1}^{d_N}  \mathcal{A}_{i_1 i_2 \cdots i_N} \mathcal{B}_{i_1 i_2 \cdots i_N} $. The Frobenius norm is defined as $|| \mathcal{A} ||_F := \sqrt{ \langle \mathcal{A}, \mathcal{A} \rangle_F } $. The spectral norm $ || \mathcal{A} ||_{\sigma} $ of the tensor $ \mathcal{ A} $ is defined as $ || \mathcal{A} ||_{\sigma} = \max \{ \mathcal{A} \otimes_1 \mathbf{x}_1 \cdots \otimes_N \mathbf{x}_N | \mathbf{x}_k \in S^{d_k - 1}, k = 1, \cdots, N \} $, where the tensor-vector product is defined as
\begin{equation*}
  \mathcal{A} \otimes_1 \mathbf{x}_1 \cdots \otimes_N \mathbf{x}_N := \sum\limits_{i_1 = 1}^{d_1} \cdots \sum\limits_{i_N = 1}^{d_N}  \mathcal{A}_{i_1 i_2 \cdots i_N} x_{1 i_1} x_{2 i_2} \cdots x_{N i_N}
\end{equation*}
and $ S^{d_k - 1}$ denotes the unit sphere in $ \mathbb{R}^{d_k} $.

\textbf{Tensor SVD $\quad$} Parallel to the matrix singular value decomposition, several orthogonal tensor decompositions with different definitions of orthogonality are studied in \cite{kolda2001orthogonal}. Among them the \emph{complete orthogonal rank decomposition} is also referred to as the \emph{tensor singular value decomposition} (tensor SVD, c.f. Definition~\ref{def:tensor_svd}) studied in \cite{chen2009tensor}. Especially, \cite{zhang2001rank} shows that for all tensors with $N \geq 3$, the tensor SVD can be uniquely determined via incremental rank-$1$ approximation.

\begin{my_definition}
 If a tensor $ \mathcal{A} \in \mathbb{R}^{d_1 \times d_2 \times \cdots \times d_N} $ can be written as sum of rank-$1$ outer product tensors $ \mathcal{A} = \sum_{i=1}^R \sigma_i u_1^{(i)} \otimes u_2^{(i)} \cdots \otimes u_N^{(i)}$, with singular values $\sigma_1 \geq \sigma_2 \geq \cdots \geq \sigma_R $ and $ \langle u_k^{(i)}, u_k^{(j)} \rangle = \delta_{ij} $ for $k=1, \cdots, N$. Then $\mathcal{A}$ has a tensor singular value decomposition with rank $R$.
 \label{def:tensor_svd}
\end{my_definition}
Define the orthogonal matrices $U_k = [ u_k^{(1)}, u_k^{(2)}, \cdots, u_k^{(R)} ] \in \mathbb{R}^{d_k \times R} $ with $ U_k^{T} U_k = \mathbb{I}_R$ for $k = 1, \cdots, N$, and the diagonal tensor $\mathcal{D} \in \mathbb{R}^{R \times R \times \cdots \times R} $ with $\mathcal{D}_{ii \cdots i} = \sigma_i$, then the tensor SVD for $\mathcal{A}$ can be also written as $\mathcal{A} = \mathcal{D} \otimes_1 U_1 \otimes_2 U_2 \cdots \otimes_N U_N$. Given an arbitrary tensor $\mathcal{A} \in \mathbb{R}^{d_1 \times d_2 \times \cdots \times d_N}$, an interesting question is to find a low-rank approximation via tensor SVD. In particular, \cite{chen2009tensor} proves the existence of the global optima of the following optimization problem
\begin{equation*}
  \min || \mathcal{A} - \sum_{i=1}^r \sigma_i u_1^{(i)} \otimes u_2^{(i)} \cdots \otimes u_N^{(i)} ||_F \ ; \text{s.t.} \ \langle u_k^{(i)}, u_k^{(j)} \rangle = \delta_{ij}, \ \text{for} \ k = 1, \cdots, N
\end{equation*}
for any $r \leq \min \{ d_1, d_2, \cdots, d_N \} $. We will utilize this fact to derive the error bound after projecting the tensor onto low-rank subspaces. Note that, in contrast to the matrix SVD, tensor SVD is unique up to the signs of singular values.


Our quantum algorithm builds on the assumption that the semantic tensor $\chi$ can be well approximated by a low-rank tensor $\hat{\chi}$ with $ || \chi - \hat{\chi} ||_F \leq \epsilon || \chi ||_F $ for small $\epsilon > 0$. Previous work of recommendation systems~\cite{drineas2002competitive} has shown that the quality of recommendations for users depends on the reconstruction error. Similarly, in the case of relational learning, with a bounded tensor approximation error it is possible to estimate the probability of a \emph{successful} information retrieval. Consider the query $(\mathrm{s}, \mathrm{p}, ?)$ (i.e., given a subject $\mathrm{s}$ and a predicate $\mathrm{p}$, to find 
the corresponding object $\mathrm{o}$) on a KG using classical algorithm. We normally only readout top-$n$ returns from the reconstructed tensor $\hat{\chi}$, written as $ \hat{x}_{\mathrm{sp} 1}, \dots, \hat{x}_{\mathrm{sp} n} $,
where $n$ is a small integer corresponding to the commonly used Hits@$n$ metric. The information retrieval is called \emph{successful} if the correct object corresponding to the query can be found in the returned list $ \hat{x}_{\mathrm{sp}1}, \dots, \hat{x}_{\mathrm{sp} n}$. In particular, we have the following estimation.

\begin{my_lemma}
If an algorithm returns an approximation of the binary semantic tensor $\chi$, denoted $\hat{\chi}$, with $ || \chi - \hat{\chi} ||_F \leq \epsilon || \chi ||_F$ and $\epsilon < \frac{1}{2}$,  then the probability of a successful information retrieval from the top-$n$ returns of $\hat{\chi}$ is at least  $1 - ( \frac{ \epsilon }{ 1 - \epsilon })^n $. (Proof in Appendix A.1)
\end{my_lemma}

In real-world applications, we can only observe part of the non-zero entries in a given tensor $\mathcal{A}$, and the task is to infer unobserved non-zero entries with high probability. This task corresponds to items recommendation for users given an observed preference matrix, or implicit knowledge inference given partially observed relational data. The partially observed tensor is called as subsampled or sparsified, denoted $\hat{ \mathcal{A} }$. Without further specifying the dimensionality of the tensor, we consider the following subsampling and rescaling scheme proposed in~\cite{achlioptas2007fast}:
\begin{equation}
  \hat{ \mathcal{A} }_{ i_1 i_2 \cdots i_N } =
  \begin{cases}
    \frac{ \mathcal{A}_{ i_1 i_2 \cdots i_N } }{ p }  & \quad  \text{with probability} \ p   \\
    0   & \quad \text{otherwise}.
  \end{cases}
  \label{eq:subsample_method_main}
\end{equation}
It means that the non-zero elements of a tensor are independently and identically sampled with the probability $p$ and rescaled afterwards. The subsampled tensor can be rewritten as $\hat{ \mathcal{A} } = \mathcal{A} + \mathcal{N}$, where $\mathcal{N}$ is a noise tensor. Entries of $\mathcal{N}$ are independent random variables with distribution $ \Pr ( \mathcal{N}_{i_1 \cdots i_N} = ( 1 / p - 1) \mathcal{A}_{i_1 \cdots i_N} ) = p  $ and $ \Pr ( \mathcal{N}_{i_1 \cdots i_N} = - \mathcal{A}_{i_1 \cdots i_N} ) = 1 - p $.

Now, the task is to reconstruct the original tensor $\mathcal{A}$ by modeling $\hat{ \mathcal{A} }$. We use tensor SVD to model the observed tensor $ \hat{\mathcal{A}} $. The reconstruction error can be bounded either using the truncated $r$-rank tensor SVD, denoted $ \hat{ \mathcal{A} }_r$, or the projected tensor SVD with absolute singular value threshold $\tau$, denoted $ \hat{ \mathcal{A} }_{ |\cdot| \geq \tau }$. Notation $ \hat{ \mathcal{A} }_{ |\cdot| \geq \tau }$ means that the subsampled tensor $ \hat{ \mathcal{A} } $ is projected onto the eigenspaces with absolute singular values larger than a cutoff threshold $\tau > 0 $. By comparison, in matrix SVD, essentially the singular values larger than, or equal to, a cutoff threshold are kept and those that are smaller are disregarded. However, in the tensor case, negative singular values can arise. The same cutoff scheme then is no longer meaningful, as it would disregard singular values with large negative values which may potentially be important.

Theorem~\ref{theorem:bound_by_rank_main} gives the reconstruction error bound using $ \mathcal{A}_r $ and the corresponding conditions on the sample probability.
\begin{my_theorem}
  Let $ \mathcal{A} \in \{ 0, 1 \}^{ d_1 \times d_2 \times \cdots \times d_N}$. Suppose that $\mathcal{A}$ can be well approximated by its $r$-rank tensor SVD $\mathcal{A}_r$. Using the subsampling scheme defined in Eq.~\ref{eq:subsample_method_main} with the sample probability $ p \geq \max \{ 0.22,  8 r \left( \log ( \frac{ 2N }{ N_0 } ) \sum\limits_{k=1}^N d_k  +\log \frac{2}{\delta} \right) /  (  \tilde{\epsilon} || \mathcal{A} ||_F )^2 \} $, $ N_0 = \log \frac{3}{2}$, then the original tensor $\mathcal{A}$ can be reconstructed from the truncated tensor SVD of the subsampled tensor $ \hat{ \mathcal{A} } $. The error satisfies $ || \mathcal{A} - \hat{ \mathcal{A}}_r ||_F  \leq  \epsilon || \mathcal{A} ||_F $ with probability at least $1 - \delta$, where $\epsilon$ is a function of $\tilde{ \epsilon }$. Especially, $\tilde{ \epsilon }$ together with the sample probability controls the norm of the noise tensor.
  \label{theorem:bound_by_rank_main}
\end{my_theorem}

\begin{proof}
  We outline the ideas involved in the proof and relegate details to the appendix A.2. The proof is divided into two parts. We first derive the following bound for the reconstruction error (see appendix Lemma A 2, 3, 4)
  \begin{equation}
    || \mathcal{A} - \hat{ \mathcal{A} }_r ||_F \leq 2 || \mathcal{A} - \mathcal{A}_r ||_F + 2 \sqrt{ || \mathcal{A}_r ||_F || \mathcal{A} - \mathcal{A}_r ||_F } \nonumber + 2 \sqrt{ || \mathcal{N}_r ||_F || \mathcal{A}_r ||_F } + || \mathcal{N}_r ||_F.
    \label{eq:reconstruction_error_main}
  \end{equation}
Notice that the RHS doesn't contain the subsampled tensor $ \hat{\mathcal{A}} $. Therefore we can further simplify the RHS by assuming that the original tensor has a low-rank approximation, namely $ || \mathcal{A} - \mathcal{A}_r ||_F \leq \epsilon_0 || \mathcal{A} ||_F $. After that, we prove numerically that the random variables $ \mathcal{ N }_{i_1 \cdots i_N} x_{1 i_1} \cdots x_{N i_N } $ for any $ \mathbf{x}_k \in S^{d_k - 1} $, $k= 1, \cdots, N$ are sub-Gaussian distributed if the sample probability fulfills $ p \gtrsim 0.22 $. Hence we can further use the covering number on the product space $ S^{d_1 - 1} \times \cdots \times S^{d_N - 1} $ to bound the norm of $ \mathcal{N} $ (see appendix Lemma A 5, 6, 7):
\begin{equation}
  || \mathcal{N}_r ||_F \leq  \sqrt{ r \frac{8}{p} \left( \log ( \frac{ 2N }{ N_0 } ) \sum\limits_{k=1}^N d_k  +\log \frac{2}{\delta} \right) }.
  \label{eq:norm_low_rank_noise}
\end{equation}
Finally, by requiring $ || \mathcal{N}_r ||_F \leq \tilde{\epsilon} || \mathcal{A} ||_F $ or by selecting
\begin{equation*}
  p \geq \max \{ 0.22,  8 r \left( \log ( \frac{ 2N }{ N_0 } ) \sum\limits_{k=1}^N d_k  +\log \frac{2}{\delta} \right) /  (  \tilde{\epsilon} || \mathcal{A} ||_F )^2 \} 
\end{equation*}
we have $ || \mathcal{A} - \hat{ \mathcal{A}}_r ||_F  \leq  \epsilon || \mathcal{A} ||_F $ via Eq.~\ref{eq:reconstruction_error_main}, where $ \epsilon := 2 ( \epsilon_0 + \sqrt{ \epsilon_0 } +  \sqrt{ \tilde{ \epsilon } } )  + \tilde{ \epsilon }$.
\end{proof}

We further introduce the projected tensor SVD $ \hat{ \mathcal{A} }_{ |\cdot| \geq \tau }$ and analyze its error bound for the later use in the quantum singular value projection. Note that quantum algorithms are fundamentally different from classical algorithms. For example, classical algorithms for matrix factorization approximate a low-rank matrix by projecting it onto a subspace spanned by the eigenspaces possessing top-$r$ singular values with predefined small $r$. Quantum subroutine, e.g., quantum singular value estimation, on the other hand, can read and store all singular values of a unitary operator into a quantum register. However, singular values stored in the quantum register cannot be read out and compared simultaneously since the quantum state collapses after one measurement; measuring the singular values one by one will also break the quantum advantage. Therefore, we perform a projection onto the union of operator's subspaces whose singular values are larger than a threshold; and this step can be implemented on the quantum register without destroying the superposition. Moreover, since we use quantum PCA as a subroutine which ignores the sign of singular values during the projection, we have to analyze the reconstruction error given by $ \hat{ \mathcal{A} }_{ |\cdot| \geq \tau }$ for the quantum algorithm. Theorem~\ref{theorem:bound_by_threshold_main} gives the reconstruction error bound using $ \hat{ \mathcal{A} }_{ |\cdot| \geq \tau } $ and conditions for the sample probability.

\begin{my_theorem}
  Let $ \mathcal{A} \in \{ 0, 1 \}^{ d_1 \times d_2 \times \cdots \times d_N}$. Suppose that $\mathcal{A}$ can be well approximated by its $r$-rank tensor SVD $\mathcal{A}_r$. Using the subsampling scheme defined in Eq.~\ref{eq:subsample_method_main} with the sample probability $ p \geq \max \{ 0.22, p_1 := \frac{ l_1 C_0 }{ ( \tilde{\epsilon} || \mathcal{A} ||_F )^2 }, p_2 := \frac{ r C_0 }{ ( \tilde{\epsilon} || \mathcal{A} ||_F )^2 }, p_3 := \frac{ \sqrt{2 r C_0 } }{ \epsilon_1 \tilde{\epsilon} || \mathcal{A} ||_F } \}$, with $\quad C_0 =  8 \left( \log ( \frac{ 2N }{ N_0 } ) \sum_{k=1}^N d_k  +\log \frac{2}{\delta} \right) $, $ N_0 =\log \frac{3}{2}$, where $l_1$ denotes the largest index of singular values of tensor $ \hat{\mathcal{A}} $ with $ \sigma_{l_1} \geq \tau $, and choosing the threshold as $ 0 < \tau \leq \frac{ \sqrt{2 C_0 }}{ p \tilde{\epsilon} }$, then the original tensor $\mathcal{A}$ can be reconstructed from the projected tensor SVD of $ \hat{ \mathcal{A} } $. The error satisfies $ || \mathcal{A} - \hat{ \mathcal{A}}_{ |\cdot| \geq \tau } ||_F  \leq  \epsilon || \mathcal{A} ||_F $ with probability at least $1 - \delta$, where $\epsilon$ is a function of $\tilde{ \epsilon }$ and $\epsilon_1$. Especially, $\tilde{ \epsilon }$ together with $p_1$ and $p_2$ determine the norm of noise tensor and $ \epsilon_1 $ together with $p_3$ control the value of $ \hat{ \mathcal{A} } $'s singular values that are located outside the projection boundary.
  \label{theorem:bound_by_threshold_main}
\end{my_theorem}
\begin{proof}
  The proof resembles that of Theorem~\ref{theorem:bound_by_rank_main}, and details are relegated in appendix A.2.  One can first derive $ || \mathcal{A} - \hat{ \mathcal{A} }_{ |\cdot| \geq \tau} ||_F \leq 3 || \mathcal{A} - \hat{ \mathcal{A} }_{l_1} ||_F $. Then we distinguish two cases: $ l_1 \geq r $ and $ l_1 < r $ and show that if $ p \geq \max \{ 0.22, p_1, p_2 \} $ it gives $ || \mathcal{N}_r ||_F \leq \tilde{\epsilon} || \mathcal{A} ||_F $ via Eq.~\ref{eq:norm_low_rank_noise}. Moreover, requiring $ p \geq p_3 $ leads to $ || \hat{ \mathcal{A} }_r - \hat{ \mathcal{A} }_{ l_1 } ||_F \leq \epsilon_1 || \mathcal{A} ||_F $. It says that the singular values of $ \hat{ \mathcal{ A } } $ that are outside the projection boundary can be controlled by $p_3$ and predefined small $ \epsilon_1 $. Notice that $p_3 \gg  {p_1, p_2}$ if tensor $ \mathcal{A}$ is dense and $ || \mathcal{A} ||_F $ is large enough. Hence we can estimate sample probability $ p \geq \{ 0.22, p_3 \} $ given predefined $ \tilde{\epsilon} $, $ \epsilon_1 $ without knowing $l_1$ a prior. On the other hand, this theorem indicates that it is impossible to complete an over sparsified tensor with subsample probability smaller than $0.22$. 
\end{proof}

In the bodies of Theorem~\ref{theorem:bound_by_rank_main} and ~\ref{theorem:bound_by_threshold_main} there exist data-dependent parameters $r$ and $l_1$ which are unknown a prior. These parameters can only be estimated by performing tensor SVD to the original and subsampled tensors explicitly. However, in practice, mostly, we are only given the subsampled tensor without even knowing the subsample probability. For example, given an incomplete semantic tensor, we do not know what percentage of information is missing, and therefore we cannot rescale the entries in the incomplete tensor. Fortunately, unlike any other matrix sparsification~\cite{achlioptas2007fast} or tensor sparsification algorithms~\cite{nguyen2010tensor}, our analysis suggests a reasonable initial guess for the subsample probability numerically, and inversely an initial guess for the lower-rank $r$ and the projection threshold $\tau$ as well.

\section{Quantum Machine Learning Algorithm for Knowledge Graphs}

\subsection{Quantum Mechanics}

To make this work self-consistent we briefly introduce the Dirac notations of quantum mechanics. Under Dirac’s convention quantum states can be represented as complex-valued vectors in a Hilbert space $\mathcal{H}$. For example, a two-dimensional complex Hilbert $\mathcal{H}_2$ space can describe the quantum state of a spin-$1$ particle, which provides the physical realization of a qubit. By default, the basis in $\mathcal{H}_2$ for a spin-$1$ qubit read $\ket{0}=[1, 0]^{\intercal}$ and $ \ket{1} = [0, 1]^{\intercal} $. The Hilbert space of a $n$-qubits system has dimension $2^n$ whose computational basis can be chosen as the canonical basis $ \ket{i} \in \{ \ket{0}, \ket{1} \}^{ \otimes n} $, where $\otimes$ represents tensor product. Hence any quantum state $\ket{\phi} \in \mathcal{H}_{ 2^n }$ can be written as a quantum superposition $ \ket{\phi} = \sum_{i=1}^{ 2^n } \phi_i \ket{i} $, where the coefficients $ |\phi_i|^2 $ can also be interpreted as the probability of observing the canonical basis state $ \ket{i} $ after measuring $\ket{ \phi }$ using canonical basis. Moreover, we use $ \bra{\phi} $ to represent the conjugate transpose of $\ket{\phi}$, i.e., $ (  \ket{ \phi } )^{ \dagger } = \bra{ \phi } $. Given two stats $ \ket{ \phi } $ and $ \ket{ \psi } $ The inner product on the Hilbert space is defined as $ \braket{ \phi | \psi }^{ \ast } = \braket{ \psi | \phi } $. A density matrix is a projection operator which is used to describe the statistics of a quantum system. For example, the density operator of the mixed state $ \ket{\phi} $ in the canonical basis reads $ \rho = \sum_{i=1}^{2^n} | \phi_i |^2 \ket{i} \bra{i} $. Moreover, given two subsystems with density matrices $\rho$ and $\sigma$ the density matrix for the whole system is their tensor product, namely $ \rho \otimes \sigma $.

The time evolution of a quantum state is generated by the Hamiltonian of the system. The Hamiltonian $H$ is a Hermitian operator with $H^{ \dagger } = H$. Let $ \ket{ \phi (t) } $ denote the quantum state at time $t$ under the evolution of an invariant Hamiltonian $H$. Then according to the Schr\"odinger equation $ \ket{ \phi (t) } = \mathrm{e}^{ - i H t} \ket{ \phi (0) } $, where the unitary operator $ \mathrm{e}^{ - i H t} $ can be written as the matrix exponentiation of the Hermitian matrix $H$, i.e., $ \mathrm{e}^{ - i H t} = \sum_{n=0}^{ \infty } \frac{ ( - i H t )^n }{ n! }$. Eigenvectors of the Hamiltonian $H$, denoted $\ket{ u_i }$, also form a basis of the Hilbert space. Then the spectral decomposition of the Hamiltonian $H$ reads $H = \sum_i \lambda_i \ket{ u_i } \bra{ u_i } $, where $\lambda_i$ is the eigenvalue or the energy level of the system. Therefore, the evolution operator of a time-invariant Hamiltonian can be rewritten as
\begin{equation}
  \mathrm{e}^{ - i H t } = \mathrm{e}^{ - i t \sum_i \lambda_i \ket{ u_i } \bra{u_i} } = \sum_i \mathrm{e}^{ - i \lambda_i t}  \ket{ u_i } \bra{ u_i },
  \label{eq:hamiltonian_evolution}
\end{equation}
where we use the observation $ ( \ket{ u_i } \bra{ u_i } )^n = \ket{ u_i } \bra{ u_i } $ for $ n = 1, \cdots, \infty $. When applying it on an arbitrary initial state $ \ket{ \phi (0) } $ we obtain $ \ket{ \phi (t) } =  \mathrm{e}^{ - i H t } \ket{ \phi (0) } = \sum_i \mathrm{e}^{ - i \lambda_i t} \beta_i \ket{ u_i } $, where $ \beta_i $ indicates the overlap between the initial state and the eigenbasis of $H$, i.e., $ \beta_i := \braket{ u_i | \phi (0) }$. To implement the time evolution operator $\mathrm{e}^{- i H t}$ and simulate the dynamics of a quantum system using universal quantum circuits is a challenging task since it involves the matrix exponentiation of a possibly dense matrix. Therefore, Hamiltonian simulation is an active research area which was first proposed by Richard Feynman~\cite{feynman1982simulating}, see also~\cite{lloyd1996universal}.


\subsection{Quantum Tensor Singular Value Decomposition }

In this section, we propose a quantum algorithm for inference on knowledge graphs using quantum singular value estimation and projection. In the following, a $3$-dimensional semantic tensor $\chi \in \{ 0, 1 \}^{ d_1 \times d_2 \times d_3}$ as one example of a tensor $\mathcal{A}$ is of particular interest. The present method builds on the assumption that the original semantic tensor $ \chi $ modeling the complete knowledge graph has a low-rank orthogonal approximation, denoted $\chi_r $, with small rank $r$. The low-rank assumption is plausible if the knowledge graph contains global and well-defined relational patterns, as has been discussed in \cite{nickel2016review}. $\chi$ could be thereof reconstructed approximately from $ \hat{ \mathcal{\chi} }$ via tensor SVD according to Theorem~\ref{theorem:bound_by_rank_main} and~\ref{theorem:bound_by_threshold_main}. Since our quantum algorithm is sampling-based instead of learning-based, w.l.o.g., we consider sampling the correct objects given the query $ (\mathrm{s}, \mathrm{p}, ?) $ as an example and discuss the runtime complexity of one inference.

Recall that the preference matrix of a recommendation system normally contains multiple nonzero entries in a given user-row; items recommendations are made according to the nonzero entries in the user-row by assuming that the user is 'typical'~\cite{drineas2002competitive}. However, in a KG there might be only one nonzero entry in the row $ ( \mathrm{s}, \mathrm{p}, \cdot) $. Therefore, we suggest, for the inference on a KG quantum algorithm needs to sample triples with the given subject $\mathrm{s}$ and post-select on the predicate $\mathrm{p}$. Post-selection can be a feasible step if the number of semantic triples with $\mathrm{s}$ as subject and $\mathrm{p}$ and predicate is $\mathcal{O}(1)$.

Before sketching the algorithm, we need to mention the quantum data structure since our method contains the preparing and exponentiating of a density matrix derived from the tensorized classical data. The most difficult technical challenges of quantum machine learning are loading classical data as quantum states and measuring the sates since reading or writing high-dimensional data from quantum states might obliterate the quantum acceleration. Therefore, the technique quantum Random Access Memory (qRAM)~\cite{giovannetti2008quantum} was developed, which can load classical data into quantum states with acceleration. Appendix A.3 gives more details on loading vector and tensorized classical data.

The basic idea of our quantum algorithm is to project the observed data onto the eigenspaces of $\hat{\chi}$ whose corresponding singular values are larger than a threshold. Therefore, we need to create an operator that can reveal the eigenspaces and singular values of $\hat{\chi}$. The first step is to prepare the following density matrix from $ \hat{\chi} $ via a tensor contraction scheme:
\begin{equation}
  \rho_{ \hat{\chi}^{\dagger} \hat{ \chi } } := \sum\limits_{i_2 i_3 i_2' i_3'} \sum\limits_{i_1} \hat{ \chi }^{ \dagger}_{ i_1, i_2 i_3 } \hat{ \chi }_{ i_1, i_2' i_3'} \ket{ i_2 i_3 } \bra{ i_2' i_3' },
\end{equation}
where $ \sum\limits_{i_1}  \hat{ \chi }^{ \dagger}_{ i_1, i_2 i_3 } \hat{ \chi }_{ i_1, i_2' i_3'}  $ means tensor contraction along the first dimension; a normalization factor is neglected temporarily. Later we will elaborate why we perform contraction along the first dimension. We have the following lemma about $\rho_{ \hat{\chi}^{\dagger} \hat{ \chi } }$ preparation.

\begin{my_lemma}
  $\rho_{ \hat{\chi}^{\dagger} \hat{ \chi } }$ can be prepared via qRAM in time $ \mathcal{O} ( \mathrm{polylog} (d_1 d_2 d_3) )$.
\end{my_lemma}
\begin{proof}
  Since $\hat{\chi} \in \mathbb{R}^{d_1 \times d_2 \times d_3} $ is a real-valued tensor, the quantum state $ \sum\limits_{ i_1 i_2 i_3 } \hat{\chi}_{ i_1 i_2 i_3 } \ket{i_1 i_2 i_3} = \sum\limits_{ i_1 i_2 i_3 } \hat{\chi}_{ i_1 i_2 i_3 } \ket{i_1} \otimes \ket{i_2} \otimes \ket{i_3} $ can be prepared via qRAM in time $ \mathcal{O} ( \mathrm{polylog} ( d_1 d_2 d_3 ) )$, where $ \ket{i_1} \otimes \ket{i_2} \otimes \ket{i_3} $ represents the tensor product of index registers in the canonical basis. The corresponding density matrix of the quantum state reads
  \begin{equation*}
    \rho =  \sum\limits_{i_1 i_2 i_3} \sum\limits_{i_1' i_2' i_3'} \hat{\chi}_{i_1 i_2 i_3} \ket{i_1} \otimes \ket{i_2} \otimes \ket{i_3} \bra{i_1'} \otimes \bra{i_2'} \otimes \bra{i_3'} \hat{\chi}_{ i_1' i_2' i_3' }^{ \dagger }.
  \end{equation*}
After preparation, a partial trace implemented on the first index register of the density matrix
  \begin{align*}
    \mathrm{tr}_1 (\rho) & =  \sum\limits_{i_2 i_3} \sum\limits_{i_2' i_3'} \sum\limits_{i_1} \hat{\chi}_{i_1 i_2 i_3} \ket{i_2} \otimes \ket{i_3} \bra{i_2'} \otimes \bra{i_3'} \hat{\chi}_{ i_1 i_2' i_3' }^{ \dagger } \\
    & = \sum\limits_{i_2 i_3 i_2' i_3'} \sum\limits_{i_1} \hat{\chi}_{i_1 i_2 i_3}^{ \dagger } \hat{\chi}_{ i_1 i_2' i_3' }  \ket{i_2 i_3} \bra{i_2' i_3'}
  \end{align*}
gives the desired operator $ \rho_{ \hat{\chi}^{\dagger} \hat{ \chi } } $.
\end{proof}
Suppose that $ \hat{ \chi } $ has a tensor SVD approximation with $ \hat{ \chi } \approx \sum_{i=1}^R \sigma_i u_1^{ (i) } \otimes u_2^{  (i) } \otimes u_3^{ (i) }$. Then the spectral decomposition of the density operator can be written as
\begin{equation*}
  \rho_{ \hat{\chi}^{\dagger} \hat{ \chi } } = \frac{1}{ \sum_{i=1}^R \sigma_i^2 } \sum\limits_{i=1}^R \sigma_i^2 \ket{u_2^{ (i) } } \otimes \ket{ u_3^{ (i) } } \bra{ u_2^{ (i) } } \otimes \bra{ u_3^{ (i) } }.
\end{equation*}
Especially, the eigenstates $ \ket{u_2^{ (i) } } \otimes \ket{ u_3^{ (i) } } $ of $ \rho_{ \hat{\chi}^{\dagger} \hat{ \chi } } $ form another set of basis in the Hilbert space of the tensor product of quantum index registers.

The next step is to readout singular values of $\rho_{ \hat{\chi}^{\dagger} \hat{ \chi } }$ and write into another quantum register via the density matrix exponentiation method proposed in~\cite{lloyd2014quantum}. This step is also referred to as quantum principal component analysis (qPCA). The key is to prepare the unitary operator
\begin{equation*}
  U := \sum\limits_{k=0}^{K-1} \ket{k \ \Delta t} \bra{k \ \Delta t}_{C} \otimes \exp (- i k \Delta t \tilde{\rho}_{ \hat{\chi}^{\dagger} \hat{ \chi } } )
\end{equation*}
which is the tensor product of a maximally mixed state $ \sum\limits_{k=0}^{K-1} \ket{k \ \Delta t} \bra{k \ \Delta t}_{C} $ with the exponentiation of the rescaled density matrix $ \tilde{\rho}_{ \hat{\chi}^{\dagger} \hat{ \chi } } $. Especially, the clock register $C$ is needed for the phase estimation and $ \Delta t$ determines the precision of estimated singular values. The following Lemma shows that the Hamiltonian simulation with unitary operator $ \mathrm{e}^{ - i t \tilde{\rho}_{ \hat{\chi}^{\dagger} \hat{ \chi } } } $ can be applied on arbitrary quantum states for any simulation time $t$.

\begin{my_lemma}
  Unitary operator $ \mathrm{e}^{ - i t \tilde{\rho}_{ \hat{\chi}^{\dagger} \hat{ \chi } } }$ can be applied to any quantum state, where $ \tilde{\rho}_{ \hat{\chi}^{\dagger} \hat{ \chi } } := \frac{ \rho_{ \hat{\chi}^{\dagger} \hat{ \chi } } }{ d_2 d_3 } $, up to simulation time $t$. The total number of steps for simulation is $ \mathcal{O} ( \frac{ t^2 }{ \epsilon } T_{\rho} )$, where $\epsilon$ is the desired accuracy, and $T_{\rho}$ is the time for accessing the density matrix.
  \label{lemma:desity_operator_evolve}
\end{my_lemma}
\begin{proof}
  The proof uses the dense matrix exponentiation method proposed in~\cite{rebentrost2018quantum}, which was developed from~\cite{lloyd1996universal}. One crucial step is to show that Hamiltonian simulation in infinitesimal time step can be implemented with a simple unitary swap operator without exponentiating the Hamiltonian. Details are in Appendix A.4 and Lemma A 8, 9.
\end{proof}

The algorithm samples triples with subject $ \mathrm{s} $ given the query $ ( \mathrm{s}, \mathrm{p}, ? ) $. Hence a quantum state $ \ket{ \hat{\chi}^{(1)}_{ \mathrm{s} } }_{ I }$ needs to be created first via qRAM in the input data register $I$, where $ \hat{\chi}^{(1)}_{ \mathrm{s} } $ denotes the $\mathrm{s}$-row of the flattened tensor $\hat{\chi}$ along the first dimension.
A flatting of a tensor $\hat{\chi} \in \mathbb{R}^{d_1 \times d_2 \times\ldots \times d_k}$ along the $i$-th ($1\le i\le k$) dimension is
\begin{align*}
    \hat{\chi}^{(i)} := \begin{bmatrix} \hat{\chi}_{1,1,\ldots,1,1,1,\ldots,1} & \hat{\chi}_{2,1,\ldots,1,1,1,\ldots,1} & \cdots & \hat{\chi}_{d_1,d_2,\ldots,d_{i-1},1,d_{i+1},\ldots,d_k} \\ \hat{\chi}_{1,1,\ldots,1,2,1,\ldots,1} & \hat{\chi}_{2,1,\ldots,1,2,1,\ldots,1} & \cdots & \hat{\chi}_{d_1,d_2,\ldots,d_{i-1},2,d_{i+1},\ldots,d_k} \\ \vdots & \vdots & & \vdots \\ \hat{\chi}_{1,1,\ldots,1,d_i,1,\ldots,1} & \hat{\chi}_{2,1,\ldots,1,d_i,1,\ldots,1} & \cdots & \hat{\chi}_{d_1,d_2,\ldots,d_{i-1},d_i,d_{i+1},\ldots,d_k} \end{bmatrix}
\end{align*}
After that, the operator $U$ is applied to the quantum state $ \sum\limits_{k=0}^{K - 1} \ket{k \Delta t}_{C} \otimes \ket{ \hat{\chi}^{(1)}_{ \mathrm{s} } }_{I} $. After this stage of computation, we obtain
\begin{equation}
  \sum\limits_{i = 1}^{R} \beta_i \left( \sum\limits_{k=0}^{K - 1} \mathrm{e}^{- i k \ \Delta t \ \tilde{\sigma}_i^2}  \ket{k \ \Delta t}_C  \right) \ket{u_i^{(2)}}_I \otimes \ket{ u_i^{(3)} }_I,
\end{equation}
where $ \tilde{\sigma}_i := \frac{ \sigma_i }{ \sqrt{ d_2d_3} }$ are the rescaled singular values of $ \tilde{\rho}_{ \hat{\chi}^{\dagger} \hat{\chi} }$ (see Eq.~\ref{eq:hamiltonian_evolution}). Moreover, $\beta_i$ are the coefficients of $\ket{ \hat{\chi}^{(1)}_{ \mathrm{s} } }_I$ decomposed in the eigenbasis $ \ket{u_2^{(i)}}_I \otimes \ket{u_3^{(i)}}_I $ of $ \rho_{ \hat{\chi}^{\dagger} \hat{ \chi } } $, namely $ \ket{ \hat{\chi}^{(1)}_{ \mathrm{s} } }_I = \sum_{i=1}^R \beta_i \ket{u_2^{(i)}}_I \otimes \ket{u_3^{(i)}}_I $.

The third step is to perform the quantum phase estimation on the clock register $C$, which is restated in the next Theorem.
\begin{my_theorem}[Quantum Phase Estimation \cite{kitaev1995quantum}]
  Let unitary $ U \ket{v_j} = \mathrm{e}^{i \theta_j} \ket{v_j} $ with $ \theta_j \in [ - \pi, \pi ]$ for $ j \in [n] $. There is a quantum algorithm that transforms $ \sum_{j \in [n]} \alpha_j \ket{v_j} \mapsto \sum_{ j \in [n] } \alpha_j \ket{v_j} \ket{ \bar{\theta}_j }$ such that $ | \bar{\theta}_j - \theta_j | \leq \epsilon $ for all $ j \in [n]$ with probability $1 - 1/ \mathrm{poly} (n)$ in time $ \mathcal{O} (T_U \log (n) / \epsilon ) $, where $T_U$ is the time to implement $U$.
  \label{theorem:phase_estimation}
\end{my_theorem}

The resulting state after phase estimation reads $ \sum_{i=1}^{R} \beta_i \ket{  \lambda_i }_C \otimes \ket{ u_2^{(i)} }_I  \otimes  \ket{ u_3^{ (i) } }_I $ where $ \lambda_i := \frac{2 \pi }{ \tilde{\sigma}_i^2 }$. In fact, it can be shown that the probability amplitude of measuring the register $C$ is maximized when $ k \ \Delta t = \lfloor \frac{ 2 \pi }{ \tilde{\sigma}_i^2 } \rceil$, where $ \lfloor \cdot \rceil $ represents the nearest integer. Therefore, the small time step $\Delta t$ determines the accuracy of quantum phase estimation. We chose $ \Delta t = \mathcal{O} ( \frac{1}{\epsilon} )$, and according to Lemma~\ref{lemma:desity_operator_evolve} the total run time is $ \mathcal{O } ( \frac{1}{\epsilon^3} T_{ \tilde{ \rho } } ) = \mathcal{O} ( \frac{1}{\epsilon^3} \mathrm{polylog} (d_1 d_2 d_3) )$. We also perform a quantum arithmetic~\cite{ruiz2017quantum} on the clock register to recover the original singular values of $ \rho_{ \hat{\chi}^{\dagger} \hat{ \chi } } $, and obtain $ \sum_{i=1}^{R} \beta_i \ket{  \sigma_i^2 }_C \otimes \ket{ u_2^{(i)} }_I  \otimes  \ket{ u_3^{ (i) } }_I $.

The next step is to perform quantum singular value projection on the quantum state obtained from the last step. Notice that, classically, this step corresponds to projecting $ \hat{\chi} $ onto the subspace $ \hat{\chi}_{ |\cdot| \geq \tau }$. In this way, observed entries will be smoothed and unobserved entries get boosted from which we can infer unobserved triples $( \mathrm{s}, \mathrm{p}, ?)$ in the test dataset (see Theorem~\ref{theorem:bound_by_threshold_main}). Quantum singular value projection given the threshold $\tau > 0$ can be implemented in the following way. We first create a new register $R$ using an auxiliary qubit and a unitary operation that maps $ \ket{\sigma_i^2}_C \otimes \ket{0}_R $ to $ \ket{\sigma_i^2}_C \otimes \ket{1}_R $ only if $ \sigma_i^2 < \tau^2 $, otherwise $ \ket{0}_R $ remains unchanged. This step of projection gives the state
\begin{equation}
   \sum\limits_{i: \sigma_i^2 \geq \tau^2 } \beta_i \ket{\sigma_i^2}_C \otimes \ket{u_2^{(i)}}_I \otimes \ket{u_3^{(i)}}_I \otimes \ket{0}_R + \sum\limits_{i: \sigma_i^2 < \tau^2 } \beta_i \ket{\sigma_i^2}_C \otimes \ket{u_2^{(i)}}_I \otimes \ket{u_3^{(i)}}_I \otimes \ket{1}_R.
   \label{eq:before_projection}
\end{equation}

\begin{algorithm}[thp]

\begin{flushleft}
\textbf{Input}:    Inference task $ ( \mathrm{s}, \mathrm{p}, ? )$  \\
\textbf{Output}:   Possible objects to the inference task
\end{flushleft}

\begin{algorithmic}[1]
\Require{Quantum access to $\hat{\chi}$ stored in a classical memory structure; threshold $\tau$ for the singular value projection}
\State{Create $ \rho_{ \hat{\chi}^{\dagger} \hat{\chi} } $ via qRAM }
\State{Create state $\ket{ \hat{\chi}_{\mathrm{s}}^{(1)} }_I $ on the input data register $I$ via qRAM}
\State{Prepare unitary operator $U$ and apply on $\ket{ \hat{\chi}_{\mathrm{s}}^{(1)} }_I $, where
  \begin{equation*}
    U := \sum\limits_{k=0}^{K - 1} \ket{k \ \Delta t} \bra{k \ \Delta t}_C \exp ( - i k \ \Delta t \ \tilde{\rho}_{ \hat{\chi}^{ \dagger } \hat{\chi} } )
  \end{equation*}}
\State{Quantum phase estimation on the clock register $C$ to obtain $ \sum_{i=1}^R \beta_i \ket{ \lambda_i }_C \otimes \ket{ u_2^{(i)} }_I \otimes \ket{ u_3^{(i)} }_I $ }
\State{Quantum arithmetic on the clock register $C$ to obtain $ \sum_{i=1}^R \beta_i \ket{ \sigma_i^2 }_C \otimes \ket{ u_2^{(i)} }_I \otimes \ket{ u_3^{(i)} }_I $}
\State{Singular value projection given the threshold $\tau$ to obtain $$ \sum\limits_{i: \sigma_i^2 \geq \tau^2 } \beta_i \ket{\sigma_i^2}_C \otimes \ket{u_2^{(i)}}_I \otimes \ket{u_3^{(i)}}_I \otimes \ket{0}_R + \sum\limits_{i: \sigma_i^2 < \tau^2 } \beta_i \ket{\sigma_i^2}_C \otimes \ket{u_2^{(i)}}_I \otimes \ket{u_3^{(i)}}_I \otimes \ket{1}_R$$ }
\State{Measure on the register $R$ and post-select the state $\ket{0}_R$ to obtain $$ \sum\limits_{i: \sigma_i^2 \geq \tau^2 } \beta_i \ket{\sigma_i^2}_C \otimes \ket{u_2^{(i)}}_I \otimes \ket{u_3^{(i)}}_I $$ }
\State{Uncompute the clock register $C$ by inversing steps 3, 4, and 5 }
\State{Measure the resulting state $  \sum\limits_{i: | \sigma_i | \geq \tau } \beta_i  \ket{u_2^{(i)}}_I \otimes \ket{u_3^{(i)}}_I $ in the canonical basis of the input register $I$}
\State{Post-select on the predicate $\mathrm{p}$ from the sampled triples $ (\mathrm{s}, \cdot, \cdot ) $}
\end{algorithmic}
\caption{Quantum Tensor SVD on Knowledge Graph}
  \label{alg:algorithm}
\end{algorithm}

The last step is to erase the clock register using reversible unitary operator $U$ again; measure the new register $R$ and post-select on the state $ \ket{0}_R$; and trace-out the clock register $C$. This leads the projected state $ \sum\limits_{i: \sigma_i^2 \geq \tau^2 } \beta_i  \ket{u_2^{(i)}}_I \otimes \ket{u_3^{(i)}}_I $. In summary, implementing all aforementioned quantum operations, in fact, produces $ \ket{ \hat{\chi}^{+}_{ |\cdot| \geq \tau } \hat{ \chi }_{ |\cdot| \geq \tau } \hat{ \chi }_{ \mathrm{s} }^{ (1) } }_I $ from the input data state $ \ket{ \hat{ \chi }_{ \mathrm{s} }^{ (1) } }_I $, where
\begin{equation*}
  \hat{\chi}^{+}_{ |\cdot| \geq \tau } \hat{ \chi }_{ |\cdot| \geq \tau } = \sum\limits_{i: | \sigma_i | \geq \tau } ( \frac{ 1 }{ \sigma_i } u_2^{(i)} \otimes u_3^{(i)} ) \otimes ( \sigma_i u_2^{(i)} \otimes u_3^{(i)} )^{ \intercal },
\end{equation*}
and $\cdot^{ + }$ represents pseudo-inverse. Now we can recover the ignored normalization factor in Eq.~\ref{eq:before_projection} and derive the probability of a successful singular value projection, which is $ \frac{ || \hat{\chi}^{+}_{ |\cdot| \geq \tau } \hat{ \chi }_{ |\cdot| \geq \tau } \hat{ \chi }_{ \mathrm{s} }^{ (1) } ||_2  }{ || \hat{ \chi }_{ \mathrm{s} }^{ (1) } ||_2 } $. Finally, we measure this state in the canonical basis to get the triples with subject $\mathrm{s}$ and post-select on the predicate $\mathrm{p}$. This will return objects to the inference $(\mathrm{s}, \mathrm{p}, ?)$ after $ \mathcal{O} ( \frac{1}{\epsilon^3} \mathrm{polylog} (d_1d_2d_3))$ times of repetitions. The quantum algorithm is summarized in Algorithm~\ref{alg:algorithm}.

\section{Experiments with Classical Tensor SVD}

At the present stage, universal quantum computers are limited by the coherence times of qubits and the fidelity for two-qubit gates. Hence, we investigate the performance of classical tensor SVD on benchmark datasets: \textsc{Kinship}, \textsc{FB15k-237}~\cite{toutanova2015observed}, and \textsc{YAGO3}~\cite{mahdisoltani2013yago3} as the verification of proposed quantum algorithm since it is essentially the quantum counterpart of classical tensor singular value decomposition method. On the other hand, the experiments can additionally verify the primary assumption that the tensor representation of a knowledge graph has a low-rank approximation if the knowledge graph contains global patterns. Table~\ref{table:statistics} provides the statistics of compared datasets. In particular, \textsc{Kinship} contains family tree relationships of two families; \textsc{FB15k-237} is a subset of Freebase which is a large collaborative knowledge base containing structured human knowledge in the triple form; \textsc{YAGO3} is a huge semantic knowledge base containing facts about persons, organizations, cities, etc.

\begin{table}[thp]
\centering
\small
  \begin{tabular}{l | c | c | c | c | c | c }
    \textbf{Dataset}   &   $\#$ Entity   &  $\#$ Predicate   &  Avg. Node Degree  &   $\#$ Train   &   $\#$ Valid   &  $\#$ Test  \\
    \hline
    \textsc{Kinship}   &  $104$  &  $25$  &  $102.8$  &  $8,544$  &  $1,068$  &  $1,074$  \\
    \textsc{FB15k-237} &  $14,541$  &  $237$  &  $21.3$  &  $272,115$  &  $17,535$  &  $20,466$  \\
    \textsc{YAGO3}     &  $123,182$  &  $37$  &  $8.8$  &  $1,079,040$  &  $5000$  &  $5000$  \\
    \hline
  \end{tabular}
  \caption{Statistics of compared datasets in the experiments, including the number of entities, predicates, average node degree, and the number of triples in the training set, validation set, and test set. Note that the average node degree is calculated as $ ( \# \text{Train} + \# \text{Valid} + \# \text{Test} ) / \# \text{Entity} $. }
  \label{table:statistics}
\end{table}
\normalsize

\begin{figure}[thp]
\centering
  \includegraphics[width=0.32\linewidth]{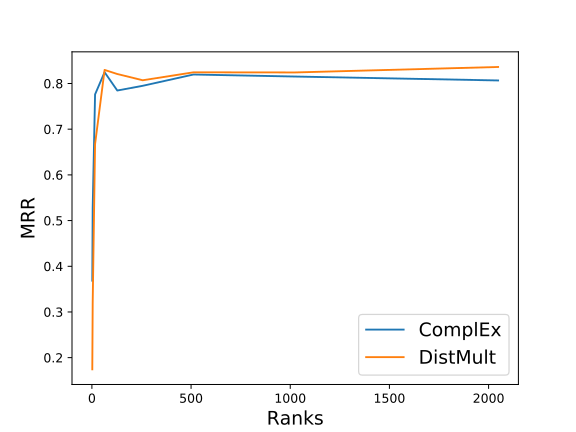}
  \includegraphics[width=0.32\linewidth]{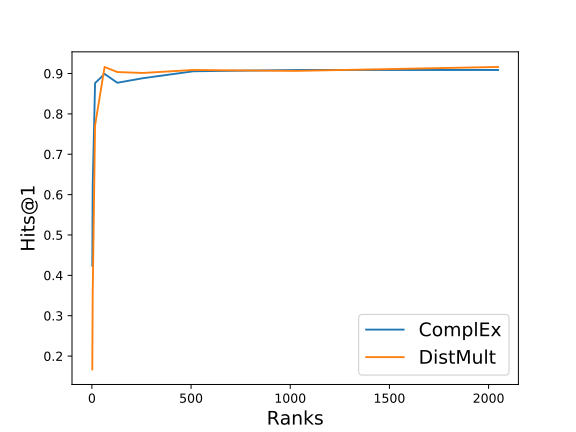}
  \includegraphics[width=0.32\linewidth]{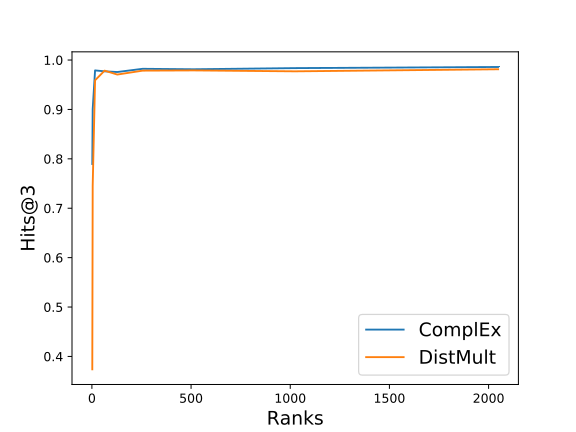}
  \caption{Low-rank representability of the \textsc{Kinship} semantic tensor by classical models \textsc{ComplEx}~\cite{trouillon2016complex} and \textsc{DistMult}~\cite{yang2014embedding}. Chosen metrics are filtered Mean Reciprocal Rank (MRR), filtered Hits@$1$, and filtered Hits@$3$, which were first introduced in~\cite{bordes2013translating} for quantifying the performance of link prediction. Metrics are evaluated on the ranks $ [2, 4, 16, 64, 128, 256, 512, 1024, 2048] $. }
  \label{fig:rank_vs_metrics}
\end{figure}

Knowledge graphs are assumed to possess structural patterns that capture global relations and rules. For instance, the composition rule $ \text{Cousin} ( X, Y ) \land \text{hasChild} (Y, Z) \rightarrow \text{Relative} (X, Z) $ implies that if $X$ is the cousin of $Y$, and $Y$ is the child of $Z$, then $X$ is a relative of $Z$. Such global relational patterns are supposed can be well learned and modeled by low-rank representation models. To experimentally verify the low-rank representability of semantic tensors, we study classical benchmark models' performance as a function of ranks on the \textsc{Kinship} dataset. We choose the \textsc{Kinship} dataset of family relationships since its relation patterns are less noisy than other datasets. The low-rank representability of the \textsc{Kinship} semantic tensor can be well observed in Figure~\ref{fig:rank_vs_metrics}, where all the metrics saturate after a small rank value.

It is, therefore, reasonable to expect that the classical counterpart of quantum tensor SVD also inherits the low-rank property. In the following, we elaborate on how to classically implement the quantum tensor SVD algorithm. Given a semantic triple $(\mathrm{s}, \mathrm{p}, \mathrm{o})$, the value function of the tensor SVD is defined as
\begin{equation*}
  \eta_{\mathrm{spo}} = \sum\limits_{i=1}^R \sigma_i \ \mathbf{u}_{ \mathrm{s} i } \ \mathbf{u}_{ \mathrm{p} i } \ \mathbf{u}_{ \mathrm{o} i }, 
\end{equation*}
where $ \mathbf{u}_{ \mathrm{s} } $, $ \mathbf{u}_{ \mathrm{p} } $, $ \mathbf{u}_{ \mathrm{o} } $ are $R$-dimensional vector representations of the subject $\mathrm{s}$, predicate $\mathrm{p}$, and object $\mathrm{o}$, respectively.  Note that the vector representations are read out from separate embedding matrices of subjects, predicates, and objects. The dimension $R$, also known as the rank, serves as a hyperparameter. In general, for benchmark classical models, we assign value $1$ to genuine semantic triples and $0$ to negative triples as ground truth labels. By formulating the link prediction task as a binary classification task, embedding matrices are learned such that the scores of genuine triples are close to $1$ and scores of negative triples close to $0$. However, for the classical counterpart of quantum tensor SVD, to reconstruct the original semantic tensor, the ground-truth values need to be rescaled by the subsample probability $p$ (see Eq.~\ref{eq:subsample_method_main}).

The model is optimized by minimizing the following objective function
\begin{equation}
  \mathcal{L} := \frac{1}{| \mathcal{D}_{\text{train}} |} \sum\limits_{(\mathrm{s, p, o}) \in \mathcal{D}_{\text{train}} } ( y_{ \mathrm{spo} } - \eta_{ \mathrm{spo} } )^2 + \gamma ( || U_s^{\intercal} U_s - \mathbb{I}_R ||_F + || U_p^{\intercal} U_p - \mathbb{I}_R ||_F + || U_o^{\intercal} U_o - \mathbb{I}_R ||_F)
  \label{eq:qsvd_loss}
\end{equation}
via stochastic gradient descent, which contains a mean square error loss and a penalization. The hyper-parameter $\gamma$ is used to encourage the orthonormality of embedding matrices for subjects, predicates, and objects as required by the definition of tensor SVD.

\begin{figure}[thp]
\centering
  \includegraphics[width=0.32\linewidth]{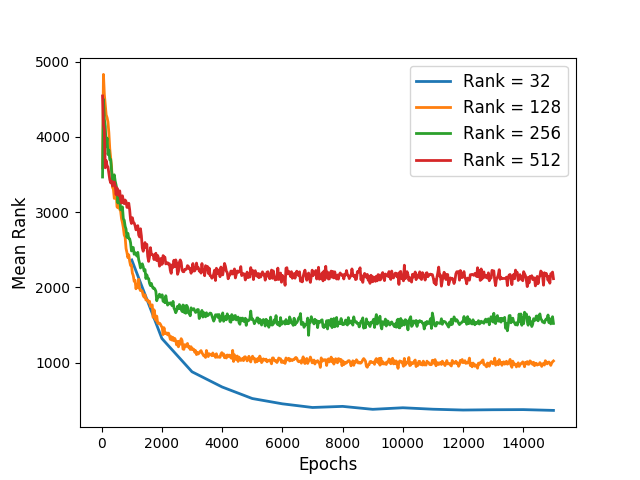}
  \includegraphics[width=0.32\linewidth]{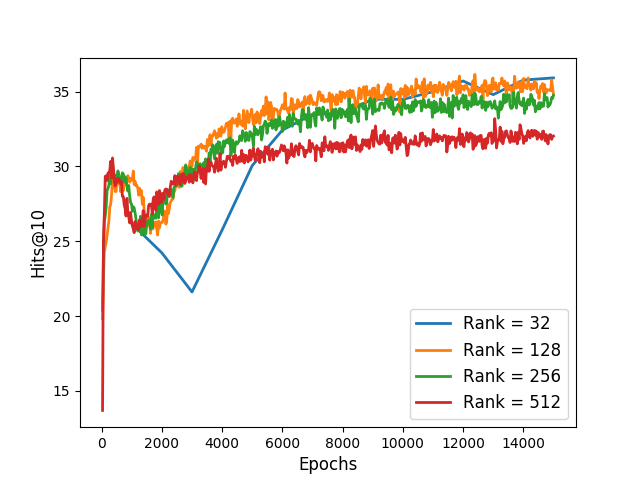}
  \includegraphics[width=0.31\linewidth]{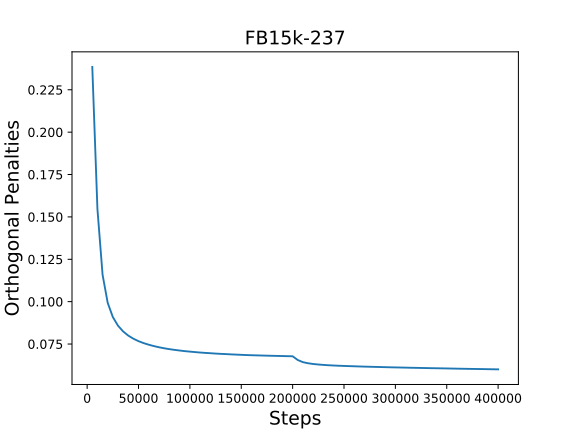}
  \caption{Learning curves of Mean Rank (left) and Hits@$10$ (middle) on the \textsc{FB15k-237} dataset for different ranks. The right panel demonstrates the decreasing orthonormality penalty of embedding matrices during the training.}
  \label{fig:rank_ortho_fb}
\end{figure}

In the left and middle panels of Figure~\ref{fig:rank_ortho_fb}, we plot training curves of the classical counterpart of quantum tensor SVD on the \textsc{FB1k-237} dataset using different rank values. The evaluation metrics are filtered Mean Rank (MR) and filtered Hits@$10$~\footnote{For the sake of simplicity, we use filtered metrics as default.} first introduced in~\cite{bordes2013translating} for quantifying the performance of link prediction. For both metrics, the classical counterpart of quantum tensor SVD performs reasonably well using only a small rank value $R=32$. This experimental observation indicates that even for the relatively noisy \textsc{FB15k-237} dataset, the corresponding semantic tensor is low-rank representable. Thus, the low-rank approximation of the tensor of the complete knowledge graph is a plausible assumption. Furthermore, knowing an approximate range of the low rank, we can estimate the projection threshold $ \tau $ necessary to the quantum algorithm according to Theorem~\ref{theorem:bound_by_threshold_main}. Besides, in the right panel of Figure~\ref{fig:rank_ortho_fb}, we show that the orthonormality penalty of embedding matrices decreases with the training.

\begin{table}[thp]
\centering
\small
  \begin{tabular}{l | c c c | c c c c | c c c c }
    &  \multicolumn{3}{c}{ \textsc{Kinship}}  &  \multicolumn{4}{c}{ \textsc{FB15k-237} }  &  \multicolumn{4}{c}{ \textsc{YAGO3}}     \\
    \textbf{Methods}  &  MR  &  @$3$  &  @$10$  &  MR  &  @$3$  &  @$10$  &  @$50$  &  MR  &  @$10$  &  @$50$  &  @$100$  \\
    \hline
    \textsc{DistMult}    &  $2.92$  &  $88.50$  &  $95.76$  &  $355.2$  &  $22.49$  &  $36.11$  &  $57.36$  &  $4841.9$  &  $46.07$  &  $65.81$  &  $\mathbf{76.28}$  \\
    \textsc{ComplEx}     &  $\mathbf{2.30}$  &  $90.32$  &  $97.77$  &  $\mathbf{343.8}$  &  $22.15$  &  $36.12$  &  $58.06$  &  $4793.7$  &  $50.10$  &  $68.83$  &  $74.35$  \\
    \textsc{RESCAL}      &  $2.40$  &  $91.06$  &  $\mathbf{98.00}$  &  $393.3$  &  $16.17$  &  $27.36$  &  $50.53$  &  $\mathbf{3589.6}$  &  $19.15$  &  $41.95$  &  $51.56$  \\
    \textsc{Tucker}      &  $2.33$  &  $\mathbf{90.78}$  &  $96.97$  &  $410.8$  &  $16.40$  &  $28.53$  &  $50.83$  &  $3590.6$  &  $24.29$  &  $45.43$  &  $54.87$  \\
    \hline
    \textsc{Tensor SVD}  &  $2.49$  &  $85.57$  &  $97.44$  &  $414.5$  &  $\mathbf{22.54}$  &  $\mathbf{37.69}$  &  $\mathbf{59.66}$  &  $6384.8$  &  $\mathbf{58.38}$  &  $\mathbf{70.90}$  &  $74.30$  \\
    \hline
    \textbf{Improve}  &  $-$  &  $-$  &  $-$  &  $-$  & $2.22\%$  &  $4.35\%$  &  $2.76\%$  &  $-$  &  $16.53\%$  & $3.00\%$  &  $-$ \\
    \hline
  \end{tabular}
  \caption{Mean Rank and Hits@$n$ metrics compared on the \textsc{Kinship}, \textsc{FB15k-237}, and \textsc{YAGO3} datasets. Relative improvements are obtained by comparing with the best classical baselines.}
  \label{table:tsvd_compare}
\end{table}
\normalsize

In Table~\ref{table:tsvd_compare}, the classical implementation of quantum tensor SVD is compared with other benchmark models, which are \textsc{RESCAL}~\cite{nickel2011three}, \textsc{Tucker}~\cite{tucker1966some}, \textsc{DistMult}~\cite{yang2014embedding}, and \textsc{ComplEx}~\cite{trouillon2016complex}. Since the magnitudes of the number of entities are different on datasets, we report different Hits@$n$ metrics on three datasets. For a fair comparison, benchmark classical models are trained using mean squared error as the loss function. Moreover, major hyperparameters for tuning the classical counterpart of quantum tensor SVD are the rank value $R$, the orthonormality penalty factor $\gamma$, and the subsample probability $p$ in Eq.~\ref{eq:subsample_method_main}. One noticeable observation is that the classical implementation of quantum tensor SVD achieves comparable or even better performance on \textsc{FB15k-237} and \textsc{YAGO3}. For the Hits@$10$ metric, the classical implementation of quantum tensor SVD even outperforms the best baseline \textsc{ComplEx} by $16.53\%$ on \textsc{YAGO3}. One possible explanation is that \textsc{YAGO3} has the lowest average node degree (see Table~\ref{table:statistics}), which implies relatively simple relational patterns in the dataset. Therefore, the quantum algorithm for sampling and reasoning on semantic knowledge graph might find important applications when the number of to be inferred entities is enormous, and when the global relational patterns are easy to learn.

\begin{figure}[thp]
\centering
  \includegraphics[width=0.32\linewidth]{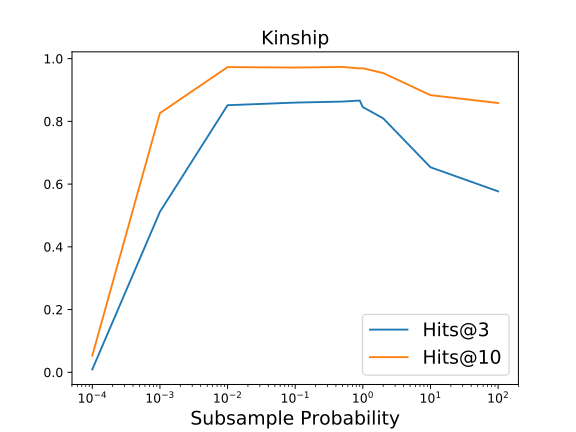}
  \includegraphics[width=0.32\linewidth]{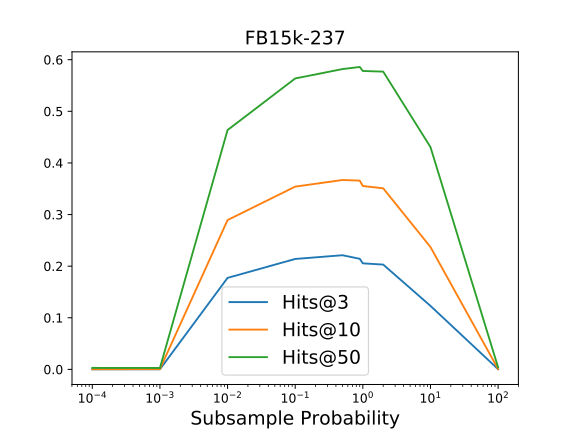}
  \includegraphics[width=0.32\linewidth]{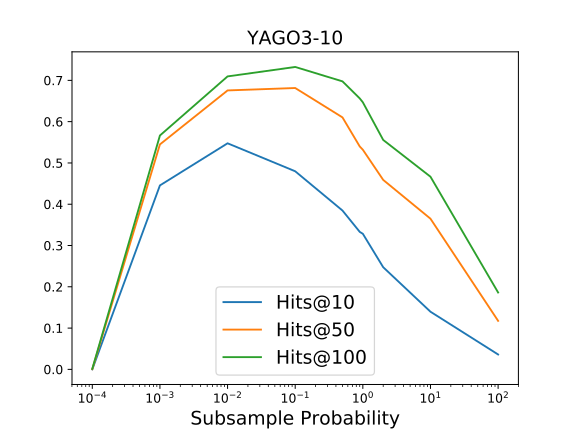}
  \caption{Hits@$n$ metrics versus the subsample probability of $p$ on datasets \textsc{Kinship}, \textsc{FB15k-237}, and \textsc{YAGO3}. Metrics are evaluated on the subsample probabilities $ p = [0.0001, 0.001, 0.01, 0.1, 0.5, 0.9, 1, 2, 10, 100] $. }
  \label{fig:metrics_vs_p}
\end{figure}

One important hyperparameter for tuning the performance of the classical counterpart of quantum tensor SVD on the validation set is the subsample probability $p$, which is used to rescale the ground-truth values before training. This hyperparameter is dataset dependent. Since, normally, $90\%$ of genuine semantic triples are randomly sampled and used as training samples, our initial guess to this hyperparameter is $p=0.9$. To verify this assumption, in Figure~\ref{fig:metrics_vs_p}, we plot Hits@$n$ metrics versus subsample probability on three datasets. One can observe that the best performance on \textsc{Kinship} and \textsc{FB15k-237} is achieved when the subsample probability $p=0.9$, which meets our expectation. However, for the \textsc{YAGO3} dataset, the best performance is found when the subsample probability equals $0.01$ or $0.1$, depending on the considered metrics. One reasonable explanation is that the \textsc{YAGO3} dataset is highly sparse, and the original dataset is already largely incomplete with a lot of missing facts.


The subsample probability is not the only factor that can affect the performance of quantum tensor SVD on knowledge graphs. Recall that given a query triple $(\mathrm{s, p, } ?)$, quantum tensor SVD returns appropriate object candidates by post-selecting the predicate $\mathrm{p}$, which is different from the classical reasoning method. To verify, in practice, whether the post-selection on predicate has an unexpected impact on the algorithm performance, we design a new task for the classical implementation of quantum tensor SVD and benchmark classical models, which is called the $(\mathrm{p, o})$-sampling test. Given a query triple $(\mathrm{s, p, o})$, we evaluate the filtered ranking of the ground-truth $(\mathrm{p, o})$ pair after calculating all the score functions whose subject is the given subject $\mathrm{s}$ in the query, namely $\eta_{\mathrm{s} \cdot \cdot }$ values.

In Table~\ref{table:po_tes}, we report filtered Hits@$n$ metrics for the novel $(\mathrm{p, o})$-sampling test on three datasets. Note that the $(\mathrm{p, o})$-sampling test is much more challenging than simply predicting the object. There are only $N_e$ candidates for the object prediction, while there are $N_p \cdot N_e$ predicate-object candidate pairs for the $(\mathrm{p, o})$ prediction, where $N_e$ is the number of entities and $N_p$ the number of predicates. Therefore, we use a higher $n$ value in the Hits@$n$ metrics for the $(\mathrm{p, o})$-sampling task. One can still observe comparable results on the largest and highly sparse \textsc{YAGO3} dataset. This observation is attractive since it reflects that even using the post-selection step, correct objects can be found in the top-$100$ returns with high probability. One can, therefore, expect that, on large and highly sparse knowledge graphs, sampling-based quantum tensor SVD algorithm might have comparable performance as benchmark classical models, while showing quantum advantages.

\begin{table}[thp]
\centering
\small
  \begin{tabular}{l | c c | c c | c c }
    &  \multicolumn{2}{c}{ \textsc{Kinship}}  &  \multicolumn{2}{c}{ \textsc{FB15k-237} }  &  \multicolumn{2}{c}{ \textsc{YAGO3}}     \\
    \textbf{Methods}  &  @$10$  &  @$50$  &  @$100$ &  @$200$  &  @$100$  &  @$200$  \\
    \hline
    \textsc{DistMult}    &  $57.4$  &  $84.8$  &  $38.0$  &  $49.5$  &  $74.0$  &  $77.0$  \\
    \textsc{ComplEx}     &  $57.0$  &  $84.8$  &  $\mathbf{39.5}$  &  $\mathbf{54.5}$  &  $\mathbf{75.0}$  &  $\mathbf{78.0}$  \\
    \textsc{RESCAL}      &  $\mathbf{75.1}$  &  $\mathbf{94.4}$  &  $15.0$  &  $27.5$  &  $62.0$  &  $69.0$  \\
    \textsc{Tucker}      &  $69.4$  &  $89.4$  &  $5.0$  &  $9.0$  &  $7.0$  &  $12.0$  \\
    \hline
    \textsc{Tensor SVD}  &  $39.6$  &  $82.0$  &  $20.0$  &  $29.5$  &  $73.5$  &  $75.0$  \\
    \hline
  \end{tabular}
  \caption{Hits@$n$ results for the novel $(\mathrm{p, o})$-sampling task on three datasets. }
  \label{table:po_tes}
\end{table}
\normalsize

\section{Conclusion}

In this work, we presented a quantum machine learning algorithm showing accelerated inference on knowledge graphs. We first proved that the semantic tensor could approximately be reconstructed from the truncated or projected tensor SVD of the subsampled tensor. Afterward, we constructed a quantum algorithm using quantum principal component analysis and singular value projection. The resulting sample-based quantum machine learning algorithm shows an acceleration w.r.t. the dimensions of the semantic tensor. Due to technical limitations, we investigate the performance of the classical counterpart of quantum tensor SVD on conventional entity prediction and novel ${\mathrm{p, o}}$-sampling tasks. The comparable, or even superior, results of the classical implementation of quantum tensor SVD on both tasks might guarantee a good performance of the quantum tensor SVD implemented on future quantum computers.

In this paper, we aim to build a first framework of quantum knowledge graph algorithms, but every component hasn't been optimized. For example, we are aware of the techniques proposed in \cite{chakraborty2019the} and \cite{gilyen2019quantum}. As future work, we will try to apply these techniques and others to improve the efficiency of our algorithm and/or avoid unnecessary computations. 


\appendix
\section{Appendix}

\subsection{Proof of Lemma 1}

\begin{appendix_lemma}[Lemma 1 in the main text]
If an algorithm returns an approximation of the binary semantic tensor $\chi$, denoted $\hat{\chi}$, with $ || \chi - \hat{\chi} ||_F \leq \epsilon || \chi ||_F$ and $\epsilon < \frac{1}{2}$,  then the probability of a successful information retrieval from the top-$n$ returns of $\hat{\chi}$ is at least  $1 - ( \frac{ \epsilon }{ 1 - \epsilon })^n $.
\end{appendix_lemma}

\begin{proof}
Since the reconstruction error of $ \chi $ from $ \hat{ \chi } $ is upper bounded, we have the following inequality
\begin{equation*}
  ( 1 - \epsilon ) || \chi ||_F \leq || \hat{ \chi } ||_F \leq ( 1 + \epsilon ) || \chi ||_F.
\end{equation*}
We can use this inequality of Frobenius norm to estimate the number of tripes which are in $ \hat{ \chi } $ but not in $ \chi $
\begin{align*}
  \epsilon^2 || \chi ||_F^2 \geq || \chi - \hat{ \chi } ||_F^2 & = \sum\limits_{ (i,j,k) \in \chi \cap (i,j,k) \in \hat{\chi} } ( 1 - \hat{ \chi}_{ijk} )^2 + \sum\limits_{ (i,j,k) \in \hat{ \chi } \cap (i, j, k) \notin \chi } \hat{ \chi }_{ijk}^2 + \sum\limits_{ (i, j, k) \in \chi \cap (i, j, k) \notin \hat{\chi} } ( 1 - \hat{ \chi }_{ijk} )^2 \\
  & \geq \sum\limits_{ (i,j,k) \in \hat{ \chi } \cap (i, j, k) \notin \chi } \hat{ \chi }_{ijk}^2,
\end{align*}
where we use the notation $ (i,j,k) \in \hat{\chi} \cap (i,j,k) \notin \chi $ to represent a semantic triple that can be observed in $ \hat{\chi } $ but not in $ \chi $, etc. Hence the probability of sampling a semantic triple from $ \hat{\chi} $ that doesn't exist in the original tensor is upper bounded by
\begin{equation*}
  \Pr [ (i,j,k) \in \hat{\chi} \cap (i,j,k) \notin \chi ] = \frac{ \sqrt{ \sum\limits_{ (i,j,k) \in \hat{ \chi } \cap (i, j, k) \notin \chi }  \hat{\chi}_{ijk}^2  } }{ || \hat{ \chi }  ||_F } \leq \frac{ \epsilon || \chi ||_F }{ || \hat{ \chi } ||_F } \leq \frac{ \epsilon }{ 1 - \epsilon }.
\end{equation*}
Without loss of generality, consider the retrieval of objects given the inference task $(\mathrm{s}, \mathrm{p}, ?)$. The retrieval becomes unsuccessful if the top-$n$ returns from $\hat{\chi}$ do not contain the correct objects regarding to the query, which has probability at most $ ( \frac{ \epsilon }{ 1 - \epsilon } )^n $. Hence the probability of a successful information retrieval from $ \hat{ \chi } $ is at least $1 - ( \frac{ \epsilon }{ 1 - \epsilon })^n $.
\end{proof}

\subsection{Proof of Theorem 1 and Theorem 2}

We first introduce and recap notations. Consider a $N$-way tensor $\mathcal{A} \in \mathbb{R}^{d_1 \times d_2 \times \cdots \times d_N}$, which has a tensor singular value decomposition with rank $R$. Let $\mathcal{A}_r = \mathcal{D} \otimes_1 U_1 \otimes_2 U_2 \cdots \otimes_N U_N$ denote the truncated $r$-rank tensor SVD of $\mathcal{A}$ with $U_i = [u_i^{(1)}, \cdots, u_i^{(r)}] \in \mathbb{R}^{d_i \times r}$ for $i=1, \cdots , N$ and $\mathcal{D} = \mathrm{diag} (\sigma_1, \cdots, \sigma_r) \in \mathbb{R}^{r \times \cdots \times r}$. Define the projection operators $\mathcal{P}_i^{\mathcal{A}, r} := \mathbb{I} \otimes \cdots \otimes U_i U_i^{T} \otimes \cdots \otimes \mathbb{I}$ with $i=1, \cdots, N$ and the product projections $\mathcal{P}^{\mathcal{A}, r} := \prod_{i=1}^N \mathcal{P}_i^{ \mathcal{A}, r }$. We have the following Lemma for the projection operator.

\begin{appendix_lemma}
  Consider a tensor $\mathcal{A}$, if $\mathcal{A}$ has an exact tensor SVD with rank $R$ then $ \mathcal{P}^{ \mathcal{A}, r} \mathcal{A} = \mathcal{A}_r $. If the tensor SVD of $\mathcal{A}$ is obtained by minimizing $ || \mathcal{A} - \sum\limits_{i=1}^R \sigma_i u_1^{(i)} \otimes u_2^{(i)} \otimes \cdots \otimes u_N^{(i)} ||_F := || \mathcal{A} - \mathcal{A}_R ||_F $, s.t. $\langle u_k^{(i)}, u_k^{(j)} \rangle = \delta_{ij}$ for $k=1, \cdots, N$ with predefined rank $R$, then $ \mathcal{P}^{ \mathcal{A}, r} \mathcal{A} = \mathcal{A}_r $ still holds.
  \label{lemma:projector}
\end{appendix_lemma}

\begin{proof}
  We first consider $\mathcal{A}$ has an exact tensor SVD. It means that $\mathcal{A} = \tilde{\mathcal{D}} \otimes_1 \tilde{U}_1 \cdots \otimes_N \tilde{U}_N$, where $ \tilde{\mathcal{D}} = \mathrm{diag} ( \sigma_1, \cdots, \sigma_R)$ and $\tilde{U}_i = [ u_i^{(1)}, u_i^{(2)}, \cdots, u_i^{(R)} ] $ for $i=1, \cdots, N$. Hence
\begin{equation*}
  \mathcal{P}^{ \mathcal{A}, r} \mathcal{A} = \tilde{ \mathcal{D} } \otimes_1 U_1 U_1^{\intercal} \tilde{U_1} \cdots \otimes_N U_N U_N^{\intercal} \tilde{U_N} = \sum\limits_{i=1}^r \sigma_i u_1^{(i)} \otimes_1 \otimes_2 u_2^{(i)} \cdots u_N^{(i)} = \mathcal{A}_r.
\end{equation*}
  On the other hand, suppose that $\mathcal{A}$'s tensor SVD is found by minimizing the objective function. Define $\mathcal{A}_R^{\perp} := \mathcal{A} - \mathcal{A}_R$, then we have $ \langle \mathcal{A}_R^{\perp} , \mathcal{T}_i \rangle = 0 $ with $ \mathcal{T}_i := u_1^{(i)} \otimes u_2^{(i)} \otimes \cdots \otimes u_N^{(i)} $ for $i = 1, \cdots, R$. To see this, suppose $\exists j$, such that $ \langle \mathcal{A}_R^{\perp}, \mathcal{T}_j \rangle = \epsilon \neq 0$. Then,
  \begin{equation*}
    || \mathcal{A} - \sum\limits_{i=1}^R \sigma_i \mathcal{T}_i - \epsilon \mathcal{T}_j ||_F^2 = || \mathcal{A} - \sum\limits_{i=1}^R \sigma_i \mathcal{T}_i ||_F^2 - \epsilon^2 < || \mathcal{A} - \sum\limits_{i=1}^R \sigma_i \mathcal{T}_i ||_F^2,
  \end{equation*}
which contradicts the fact that $\mathcal{A}_R$ is the global minimum of the objective function. Thus, $\mathcal{P}^{ \mathcal{A}, r} \mathcal{A} = \mathcal{P}^{ \mathcal{A}, r} ( \mathcal{A}_R + \mathcal{A}_R^{\perp} ) = \mathcal{P}^{ \mathcal{A}, r} \mathcal{A}_R = \mathcal{A}_r $.
\end{proof}

As we can see the projection operator $ \mathcal{P}^{\mathcal{A}, r} $ projects the tensor onto the space spanned by $\mathcal{T}_i, \cdots, \mathcal{T}_r$. Lemma A~\ref{lemma:projector} also implies that for any two tensors $\mathcal{A}$ and $\mathcal{B}$ we have the inequality
\begin{equation}
  || \mathcal{P}^{ \mathcal{A}, r} \mathcal{A} ||_F \geq || \mathcal{P}^{ \mathcal{B}, r} \mathcal{A} ||_F.
  \label{eq:projection_inequality}
\end{equation}
In the next Lemma we give the lower bound  of $ || \mathcal{P}^{ \mathcal{B}, r} \mathcal{A} ||_F $. The proof is similar to the matrix case which is given in~\cite{achlioptas2007fast}.

\begin{appendix_lemma}
  Given two tensors $\mathcal{A}$ and $\mathcal{B}$ having tensor SVD with ranks $R_A$ and $R_B$, respectively. Suppose $r \leq \min \{ R_A, R_B \}$, we have
\begin{equation*}
  || \mathcal{P}^{ \mathcal{B}, r} \mathcal{A} ||_F \geq || \mathcal{P}^{ \mathcal{A}, r} \mathcal{A} ||_F - 2 || \mathcal{P}^{ \mathcal{A} - \mathcal{B}, r } ( \mathcal{A} - \mathcal{B} ) ||_F.
\end{equation*}
  \label{lemma:bound_proj_A_on_B}
\end{appendix_lemma}
\begin{proof}
  \begin{align*}
    ||  \mathcal{P}^{ \mathcal{B}, r } \mathcal{A} ||_F & = || \mathcal{P}^{ \mathcal{B}, r} ( \mathcal{B} + ( \mathcal{A} - \mathcal{B} ) ) ||_F \geq || \mathcal{P}^{ \mathcal{B}, r } \mathcal{B} ||_F - || \mathcal{P}^{ \mathcal{B}, r } ( \mathcal{A} - \mathcal{B} ) ||_F  \\
    & \geq || \mathcal{P}^{ \mathcal{A}, r } \mathcal{B} ||_F - || \mathcal{P}^{ \mathcal{B}, r } ( \mathcal{A} - \mathcal{B} ) ||_F  =   || \mathcal{P}^{ \mathcal{A}, r } ( \mathcal{A} - ( \mathcal{A} - \mathcal{B} ) ) ||_F - || \mathcal{P}^{ \mathcal{B}, r } ( \mathcal{A} - \mathcal{B} ) ||_F \\
    & \geq || \mathcal{P}^{ \mathcal{A}, r } \mathcal{A} ||_F - || \mathcal{P}^{ \mathcal{A}, r } ( \mathcal{A} - \mathcal{B} ) ||_F - || \mathcal{P}^{ \mathcal{B}, r } ( \mathcal{A} - \mathcal{B} ) ||_F \\
    & \geq || \mathcal{P}^{ \mathcal{A}, r } \mathcal{A} ||_F - 2  || \mathcal{P}^{ \mathcal{A} - \mathcal{B}, r } ( \mathcal{A} - \mathcal{B} ) ||_F,
  \end{align*}
where we used Eq.~\ref{eq:projection_inequality} multiple times.
\end{proof}

Lemma A~\ref{lemma:bound_proj_A_on_B} indicates that if $ \mathcal{A} $ and $ \mathcal{B} $ are similar tensors, then the  projection of tensor $ \mathcal{A} $ onto the first $r$ bases of tensor $ \mathcal{B} $ has only small error which is bounded by $ || \mathcal{P}^{ \mathcal{A} - \mathcal{B}, r } ( \mathcal{A} - \mathcal{B} ) ||_F $. Using Lemma A~\ref{lemma:bound_proj_A_on_B} we can derive the following bound which will serve as the main Lemma for estimating the bound of reconstruction error.

\begin{appendix_lemma}
  Given two tensors $\mathcal{A}$ and $\mathcal{B}$ having tensor SVD with ranks $R_A$ and $R_B$, respectively. Suppose $r \leq \min \{ R_A, R_B \}$, we have
\begin{align*}
  & || \mathcal{A} - \mathcal{P}^{ \mathcal{B}, r } \mathcal{B} ||_F \\
  & \ \leq 2 || \mathcal{A} - \mathcal{A}_r ||_F + 2 \sqrt{ || \mathcal{A}_r ||_F || \mathcal{A} - \mathcal{A}_r ||_F } + 2\sqrt{ || \mathcal{A}_r ||_F || \mathcal{P}^{ \mathcal{A} - \mathcal{B}, r } ( \mathcal{A} - \mathcal{B} ) ||_F } + || \mathcal{P}^{ \mathcal{A} - \mathcal{B}, r } ( \mathcal{A} - \mathcal{B} ) ||_F.
\end{align*}
  \label{lemma:bound_A_B_r}
\end{appendix_lemma}

\begin{proof}
  \begin{align*}
    || \mathcal{A} - \mathcal{P}^{ \mathcal{B}, r } \mathcal{B} ||_F & = || \mathcal{A} - \mathcal{P}^{ \mathcal{B}, r} ( \mathcal{A} - ( \mathcal{A} - \mathcal{B}) ) ||_F \leq || \mathcal{A} - \mathcal{P}^{ \mathcal{B}, r } \mathcal{A} ||_F + || \mathcal{P}^{ \mathcal{B}, r } ( \mathcal{A} - \mathcal{B} ) ||_F \\
    & \leq || \mathcal{P}^{ \mathcal{A}, r } \mathcal{A} - \mathcal{P}^{ \mathcal{B}, r } \mathcal{A} ||_F + || \mathcal{A} - \mathcal{P}^{ \mathcal{A}, r } \mathcal{A}  ||_F  +  || \mathcal{P}^{ \mathcal{B}, r } ( \mathcal{A} - \mathcal{B} ) ||_F  \\
    & = || \mathcal{A}_r - \mathcal{P}^{ \mathcal{B}, r } ( ( \mathcal{A} - \mathcal{A}_r ) + \mathcal{A}_r ) ||_F + || \mathcal{A} - \mathcal{P}^{ \mathcal{A}, r } \mathcal{A} ||_F + || \mathcal{P}^{ \mathcal{B}, r } ( \mathcal{A} - \mathcal{B} ) ||_F \\
    & \leq || \mathcal{A}_r - \mathcal{P}^{ \mathcal{B}, r } \mathcal{A}_r ||_F + || \mathcal{P}^{ \mathcal{B}, r } ( \mathcal{A} - \mathcal{A}_r ) ||_F + || \mathcal{A} - \mathcal{P}^{ \mathcal{A}, r } \mathcal{A} ||_F + || \mathcal{P}^{ \mathcal{B}, r } ( \mathcal{A} - \mathcal{B} ) ||_F \\
    & \leq \underbrace{ || \mathcal{A}_r - \mathcal{P}^{ \mathcal{B}, r } \mathcal{A}_r ||_F }_{ (\star) } + 2 || \mathcal{A} - \mathcal{P}^{ \mathcal{A}, r } \mathcal{A} ||_F + || \mathcal{P}^{ \mathcal{A} - \mathcal{B}, r } ( \mathcal{A} - \mathcal{B} ) ||_F,
  \end{align*}
for the last inequality we use Eq.~\ref{eq:projection_inequality} multiple times. Now we can apply Pythagorean theorem on the first $r$ eigenbases of tensor $\mathcal{A}$ to bound the term $(\star)$. Hence
\begin{align*}
  (\star) & = \sqrt{ || \mathcal{A}_r ||_F^2 - || \mathcal{P}^{ \mathcal{B}, r } \mathcal{A}_r ||_F^2} \\
   & \stackrel{(1)}{ \leq } \sqrt{ || \mathcal{A}_r ||_F^2 - || \mathcal{A}_r ||_F^2 + 4 || \mathcal{A}_r ||_F || \mathcal{P}^{ \mathcal{A}_r - \mathcal{B}, r } ( \mathcal{A}_r - \mathcal{B} ) ||_F } \\
   & = 2 \sqrt{ || \mathcal{A}_r ||_F || \mathcal{P}^{ \mathcal{A}_r - \mathcal{B}, r } ( \mathcal{A}_r - \mathcal{B} ) ||_F  }  \\
   & \leq 2 \sqrt{ || \mathcal{A}_r ||_F [ || \mathcal{P}^{ \mathcal{A}_r - \mathcal{B}, r } ( \mathcal{A}_r  - \mathcal{A}) ||_F + || \mathcal{P}^{ \mathcal{A}_r - \mathcal{B}, r } ( \mathcal{A} - \mathcal{B} ) ||_F ] }  \\
   & \stackrel{ (2) }{ \leq } 2 \sqrt{ || \mathcal{A}_r ||_F [ || \mathcal{A}_r  - \mathcal{A}||_F + || \mathcal{P}^{ \mathcal{A} - \mathcal{B}, r } ( \mathcal{A} - \mathcal{B} ) ||_F ] } \\
   & \stackrel{(3)}{ \leq } 2 \sqrt{ || \mathcal{A}_r ||_F || \mathcal{A} - \mathcal{A}_r ||_F } + 2 \sqrt{ || \mathcal{A}_r ||_F || \mathcal{P}^{ \mathcal{A} - \mathcal{B}, r } ( \mathcal{A} - \mathcal{B} ) ||_F },
\end{align*}
where inequality $(1)$ is given by Lemma A~\ref{lemma:bound_proj_A_on_B}, $(2)$ by Eq.~\ref{eq:projection_inequality} and $(3)$  is according to $ \sqrt{x + y} \leq \sqrt{x} + \sqrt{y}$.
\end{proof}
In summary, we have the following bound
\begin{align*}
  & || \mathcal{A} - \mathcal{P}^{ \mathcal{B}, r } \mathcal{B} ||_F  \\
  & \ \leq 2 || \mathcal{A} - \mathcal{A}_r ||_F + 2 \sqrt{ || \mathcal{A}_r ||_F || \mathcal{A} - \mathcal{A}_r ||_F } + 2\sqrt{ || \mathcal{A}_r ||_F || \mathcal{P}^{ \mathcal{A} - \mathcal{B}, r } ( \mathcal{A} - \mathcal{B} ) ||_F } + || \mathcal{P}^{ \mathcal{A} - \mathcal{B}, r } ( \mathcal{A} - \mathcal{B} ) ||_F.
\end{align*}

Consider a tensor $\mathcal{A}$ which will be subsampled and rescaled. The resulting perturbed tensor can be written as $\hat{ \mathcal{A} } = \mathcal{A} + \mathcal{N}$, where $\mathcal{N}$ is a noise tensor. In the following, we use $\hat{ \mathcal{A} }$ to represent subsampled (sparsified) tensor, and $\hat{\mathcal{A}}_r$ the truncated $r$-rank tensor SVD of $\hat{\mathcal{A}}$. Thus, according to Lemma A~\ref{lemma:bound_A_B_r} the reconstruction error using the truncated tensor SVD of the sparsified tensor $ \hat{ \mathcal{A} } $ is upper bounded by
\begin{equation}
 || \mathcal{A} - \hat{ \mathcal{A} }_r ||_F \leq 2 || \mathcal{A} - \mathcal{A}_r ||_F + 2 \sqrt{ || \mathcal{A}_r ||_F || \mathcal{A} - \mathcal{A}_r ||_F } + 2 \sqrt{ || \mathcal{N}_r ||_F || \mathcal{A}_r ||_F } + || \mathcal{N}_r ||_F.
  \label{eq:reconstruction_error}
\end{equation}

To further estimate the bound of the error, we briefly recap the tensor subsampling and sparsification techniques. The basic idea behind matrix/tensor sparsification algorithms is to neglect all small entries and keep or amplify sufficiently large entries, such that the original matrix/tensor can be reconstructed element-wise with bounded error. Matrix sparsification was first studied in~\cite{achlioptas2007fast}, and tensor sparsification in~\cite{nguyen2010tensor}.

Without further specification, we consider the following general sparsification and rescaling method used in the main text:
\begin{equation}
  \hat{\mathcal{A}}_{i_1 i_2 \cdots i_N} =
    \begin{cases}
      \frac{\mathcal{A}_{i_1 i_2 \cdots i_N}}{p}   \quad & \text{with probability} \ p > 0  \\
      0 & \text{otherwise},
    \end{cases}
  \label{eq:subsample_method}
\end{equation}
where the choose of the element-wise sample probability $p$ will be discussed later. Note that the expectation values of the entries of the sparsified tensor read $\mathbb{E} [ \hat{\mathcal{A}}_{i_1 i_2 \cdots i_N} ] = \mathcal{A}_{i_1 i_2 \cdots i_N}$. Recall that the perturbation is defined as $\mathcal{N} = \hat{\mathcal{A}} - \mathcal{A}$. Thus, the entries of the noise tensor have zero mean $\mathbb{E}[ \mathcal{N}_{i_1 i_2 \cdots i_N} ] = 0$ and variance $ \mathrm{Var} [ \mathcal{N}_{i_1 i_2 \cdots i_N} ] = \mathcal{A}_{i_1 i_2 \cdots i_N}^2 ( \frac{1}{p} - 1) $.

To bound the norms of the noise tensor $\mathcal{N}$ we also need the following auxiliary lemmas.

\begin{figure}[htp]
\begin{center}
  \begin{tabular}{c c}
    \hspace{-1.8cm}
    \includegraphics[width=0.65\linewidth]{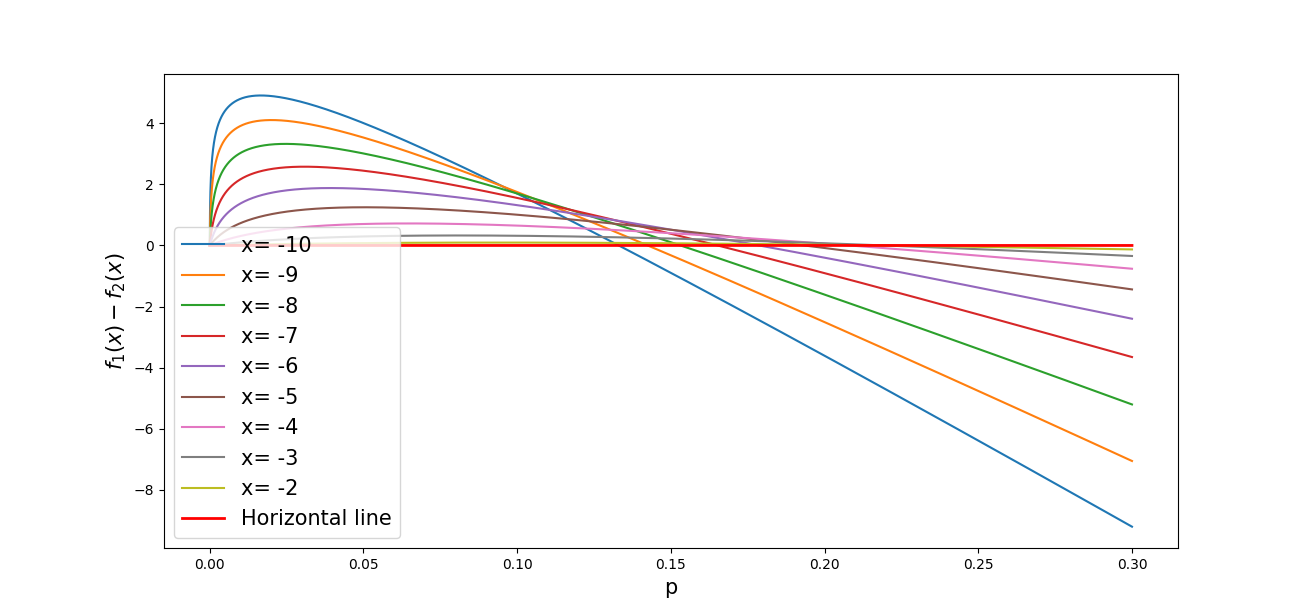}
    \hspace{-1cm}
    \includegraphics[width=0.65\linewidth]{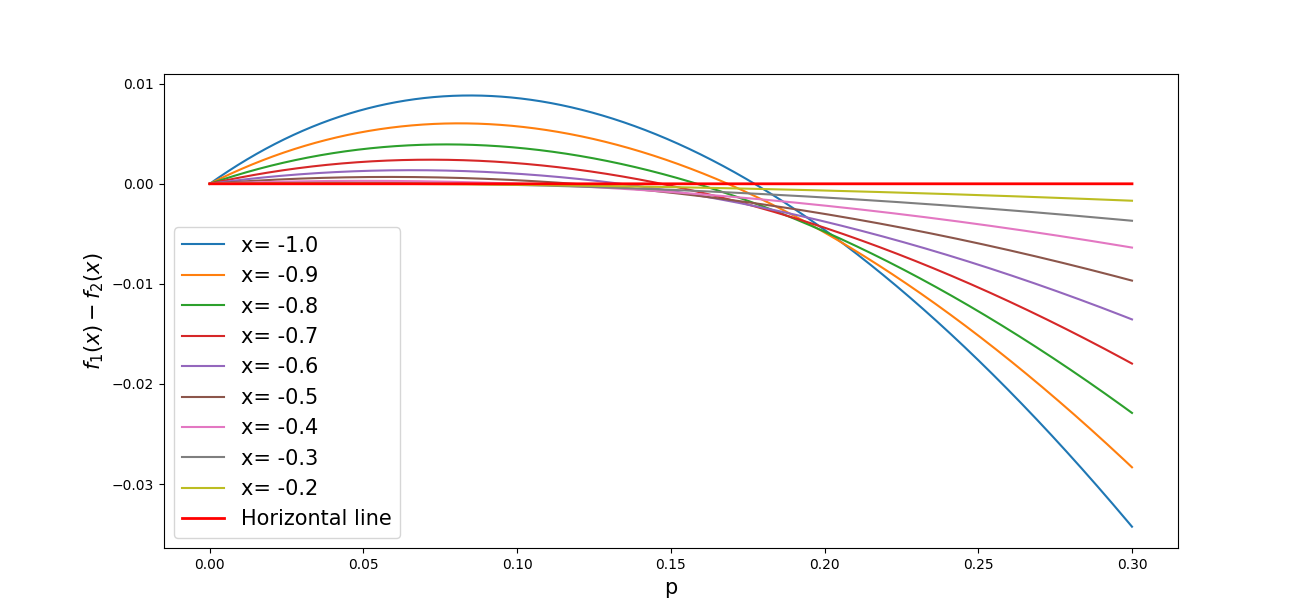}
  \end{tabular}
  \caption{Plotting $f_1(x, p) - f_2(x, p)$ for $x=[-10, -9, -8, -7, -6, -5, -4, -3, -2]$ (left) and $x=[-1., -0.9, -0.8, -0.7, -0.6, -0.5, -0.4, -0.3, -0.2]$ (right).}
  \label{fig:p_vs_diff}
\end{center}
\end{figure}

\begin{appendix_lemma}
  Define two functions $f_1(x) = px + \ln (1 - p + p \ \mathrm{e}^{-x})$ and $ f_2(x) = px^2 / 2 $. For any $x \in ( - \infty ,  \infty ) $ and $0.22 \leq p \leq 1$, we have $f_1(x) \leq f_2(x)$.
 \label{lemma:compare}
\end{appendix_lemma}
\begin{proof}
  We first consider the case when $x \geq 0$. First we have $f_1(0) = f_2(0)$ and $f_1'(0) = f_2'(0)$. Since
\begin{align*}
  1 - p + p \ \mathrm{e}^{-x} & = (\sqrt{1-p} - \sqrt{\mathrm{e}^{-x}})^2 + 2 \sqrt{(1-p) \mathrm{e}^{-x}} - \mathrm{e}^{-x} + p \ \mathrm{e}^{-x} \\
  & \geq 2 \sqrt{(1-p) \mathrm{e}^{-x} } - (1-p) \mathrm{e}^{-x},
\end{align*}
we immediately have the following inequality for the second derivatives of $f_1(x)$ and $f_2(x)$,
\begin{align}
  f_1''(x) & = \frac{ p(1-p) \mathrm{e}^{ -x } }{ (1 - p + p \ \mathrm{e}^{-x})^2 } \leq \frac{ p (1-p) \mathrm{e}^{-x} }{ ( 2 \sqrt{ (1-p) \mathrm{e}^{-x} } - (1-p) \mathrm{e}^{-x} )^2 }   \nonumber  \\
  & \leq \frac{ p (1-p) \mathrm{e}^{-x} }{ ( \sqrt{ (1-p) \mathrm{e}^{-x} } )^2 } = p = f_2''(x).
  \label{eq:2nd_ineq}
\end{align}
We used the condition that $0 \leq p \leq 1$ and $\mathrm{e}^{-x} \leq 1$ for $x \geq 0$ to derive the second inequality in Eq.~\ref{eq:2nd_ineq}. Hence $f_1(x) \leq f_2(x)$ for any $x \geq 0$ and $0 \leq p \leq 1$.

  Next, we consider the case when $ x < 0$ for different values of $p$. To find the condition of non-negative $p$ such that $f_1(x) \leq f_2(x)$ we need to solve a transcendent inequality numerically. Hence in Figure~\ref{fig:p_vs_diff} we plot $f_1(x) - f_2(x)$ as a function of $x$ and $p$. From Figure~\ref{fig:p_vs_diff} we can read the following numerical conditions
\begin{table}[htp]
\footnotesize
\centering
\begin{tabular}{c | c || c | c || c | c || c | c }
  \hline
  $x=0.$    &  $ p \geq 0. $      & $x=-0.5$ & $ p \geq 0.1185$ & $x=-1$ & $p \geq 0.1772$ & $x=-6$ & $p \geq 0.1787$ \\
  \hline
  $x=-0.1$  &  $ p \geq 0.0310 $  & $x=-0.6$ & $ p \geq 0.1337$ & $x=-2$ & $p \geq 0.2184$ & $x=-7$ & $p \geq 0.1652$ \\
  \hline
  $x=-0.2$ & $p \geq 0.0577$ & $x=-0.7$ & $p \geq 0.1469$ & $x=-3$ & $p \geq \mathbf{0.2196} $ & $x=-8$ & $p \geq 0.1531$ \\
  \hline
  $x=-0.3$ & $p \geq 0.0809$ & $x=-0.8$ & $p \geq 0.1585$ & $x=-4$ & $p \geq 0.2082$ & $x=-10$ & $p \geq 0.1331$ \\
  \hline
  $x=-0.4$ & $p \geq 0.1010$ & $x=-0.9$ & $p \geq 0.1685$ & $x=-5$ & $p \geq 0.1934$ & $x=-20$ & $p \geq 0.0794$ \\
  \hline
\end{tabular}
\caption{Numerical conditions for non-negative $p$ such that $f_1(x, p) - f_2(x, p) \leq 0$ for different values of $x$.}
\label{table:numerical_conditions}
\end{table}
\normalsize

Table~\ref{table:numerical_conditions} and Figure~\ref{fig:p_vs_diff} indicate that if $p \gtrsim 0.22$ we have $f_1(x) \leq f_2(x)$ for any $ x < 0 $ in the worst case. In summary, $f_1(x) \leq f_2(x)$ for any $x \in ( - \infty,  \infty ) $ and $0.22 \leq p \leq 1$.
\end{proof}

\begin{appendix_lemma}
  Assume that the noise tensor $\mathcal{N}$ is generated by subsampling a binary tensor $\mathcal{A} \in \{ 0, 1 \}^{d_1 \times d_2 \times \cdots \times d_N}$ according to Eq.~\ref{eq:subsample_method} with sample probability $p \gtrsim 0.22 $. The spectral norm of $\mathcal{N}$ is bounded by
  \begin{equation}
    || \mathcal{N} ||_{ \sigma } \leq \sqrt{ \frac{8}{p} \left( \log ( \frac{ 2N }{ N_0 } ) \sum\limits_{k=1}^N d_k  +\log \frac{2}{\delta} \right)  },
  \end{equation}
with probability at least $1 - \delta$.
\end{appendix_lemma}

\begin{proof}
Recall that the noise tensor entries $\mathcal{N}_{i_1 i_2 \cdots i_N}$ are independent random variables with zero mean and
  \begin{equation*}
    \mathcal{N}_{i_1 i_2 \cdots i_N} =
    \begin{cases}
      ( \frac{1}{p} - 1) \mathcal{A}_{i_1 i_2 \cdots i_N}  \quad & \text{with probability} \ p  \\
      - \mathcal{A}_{i_1 i_2 \cdots i_N}  \quad & \text{with probability} \ 1 - p.
    \end{cases}
  \end{equation*}

We first estimate the quantity $\mathbb{E} [ \mathrm{e}^{ -t \mathcal{N}_{i_1 i_2 \cdots i_N} x_{1 i_1} x_{2 i_2} \cdots x_{N i_N} } ]$ for any $t \geq 0$ with $\mathbf{x}_k \in S^{d_k - 1}$, $k=1, \cdots, N$. For the sake of succinct notation we adopt a bijection of index and write $\mathcal{N}_l := \mathcal{N}_{i_1 i_2 \cdots i_N}$ and $x_l := x_{1 i_1} x_{2 i_2} \cdots x_{N i_N} $ for $l = 1, \cdots, d_1 d_2 \cdots d_N$. Then we have the following inequality via Lemma A~\ref{lemma:compare}
\begin{align*}
  \mathbb{E} [ \mathrm{e}^{ -t \mathcal{N}_l x_l } ] & = p \ \mathrm{e}^{ -t ( \frac{1}{p} - 1 ) \mathcal{A}_l x_l } + (1-p) \ \mathrm{e}^{ t \mathcal{A}_l x_l } = \mathrm{e}^{t \mathcal{A}_l x_l } \left( 1 - p + p \ \mathrm{e}^{ - \frac{t}{p} \mathcal{A}_l x_l } \right) \\
  & = \mathrm{e}^{ p y + \ln ( 1 - p + p \mathrm{e}^{-y}) } \leq \mathrm{e}^{ \frac{p y^2 }{ 2 } } \quad \text{for} \quad p \gtrsim 0.22,
\end{align*}
where $ y:= \frac{ t \mathrm{A}_l x_l }{ p }$. Since $\mathcal{A}_l \in [0, 1]$, we have $ \mathbb{E} [ \mathrm{e}^{ -t \mathcal{N}_l x_l } ] \leq \mathrm{e}^{ \frac{t^2}{2 p} x_l^2 } $ for any $t \geq 0 $. In other words, random variables $ \mathcal{N}_l x_l $ are sub-Gaussian distributed if the sample probability fulfills $ p \gtrsim 0.22 $.

Hence
\begin{align*}
  \mathbb{E} [ \mathrm{e}^{-t \sum_l \mathcal{N}_l x_l } ] & = \mathbb{E} [ \mathrm{e}^{ -t \mathcal{N} \otimes_1 \mathbf{x}_1 \cdots \otimes_N \mathbf{x}_N } ] \leq \prod_l \mathrm{e}^{ \frac{t^2}{2p} x_l^2 }  \\
  & = \mathrm{e}^{ \frac{ t^2 }{ 2p } \sum_{i_1=1}^{d_1} x_{1 i_1}^2 \sum_{i_2=1}^{d_2} x_{2 i_2}^2 \cdots \sum_{i_N=1}^{d_N} x_{N i_N}^2} = \mathrm{e}^{ \frac{t^2}{ 2p } },
\end{align*}
where we use $|| \mathbf{x}_k ||_2 = 1$, $k=1, \cdots, N$.

Given non-negative auxiliary parameters $\lambda$ and $t$, we have
\begin{align*}
  \Pr ( \mathcal{N} \otimes_1 \mathbf{x}_1 \cdots \otimes_N \mathbf{x}_N \leq - \lambda ) & = \Pr ( \mathrm{e}^{ - t \mathcal{N} \otimes_1 \mathbf{x}_1 \cdots \otimes_N \mathbf{x}_N } \geq \mathrm{e}^{ t \lambda } ) \\
  & \leq \mathrm{e}^{ - t \lambda } \mathbb{E} [ \mathrm{e}^{ - t \mathcal{N} \otimes_1 \mathbf{x}_1 \cdots \otimes_N \mathbf{x}_N } ] \\
  & \leq \mathrm{e}^{ \frac{t^2}{2p} - t \lambda } \leq \mathrm{e}^{ - \frac{ p \lambda^2 }{2} }
\end{align*}
by choosing $t = p\lambda$. Similarly we have the probability $ \Pr ( \mathcal{N} \otimes_1 \mathbf{x}_1 \cdots \otimes_N \mathbf{x}_N \geq \lambda ) \leq \mathrm{e}^{ - \frac{ p \lambda^2 }{ 2 } } $. In summary,
\begin{equation}
   \Pr ( | \mathcal{N} \otimes_1 \mathbf{x}_1 \cdots \otimes_N \mathbf{x}_N | \geq \lambda ) \leq  2 \mathrm{e}^{ - \frac{ p \lambda^2 }{ 2 } },
   \label{eq:auxiliary_bound}
\end{equation}
if $ \mathbf{x}_k \in S^{d_k - 1}$, $k=1, \cdots, N$ and $p \geq 0.22$.

Now we are able to use the covering number argument proposed in~\cite{tomioka2014spectral} to bound the spectral norm. Let $\mathcal{C}_1, \cdots, \mathcal{C}_N$ be the $\epsilon$-covering of spheres $S^{d_1-1}, \cdots, S^{d_N - 1}$ with covering number $| \mathcal{C}_k |$ upper bounded by $ ( \frac{2}{ \epsilon } )^{ d_k } $ for $k = 1, \cdots, N$. Since the product space $ S^{d_1-1} \times \cdots \times S^{d_N-1} $ is closed and bounded, there is a point $( \mathbf{x}_1^{\star}, \cdots, \mathbf{x}_N^{\star} ) \in S^{d_1-1} \times \cdots \times S^{d_N-1} $ which maximizes the tensor-vector product $ \mathcal{N} \otimes_1 \mathbf{x}_1 \cdots \otimes_N \mathbf{x}_N $. Hence
\begin{equation}
  || \mathcal{N} ||_{ \sigma } = \mathcal{N} \otimes_1 ( \bar{\mathbf{x}}_1 + \boldsymbol{ \delta }_1 ) \cdots \otimes_N ( \bar{\mathbf{x}}_N + \boldsymbol{ \delta }_N ),
  \label{eq:write_norm_in_covering}
\end{equation}
where $ \bar{\mathbf{x}}_k + \boldsymbol{ \delta }_k = \mathbf{x}_k^{\star} $ and $ \bar{\mathbf{x}}_k \in \mathcal{C}_k $ for $k = 1, \cdots, N $. According to the definition of $\epsilon$-covering, we have $ || \boldsymbol{ \delta }_k ||_2 \leq \epsilon $.

Expanding Eq.~\ref{eq:write_norm_in_covering} gives
\begin{equation*}
  || \mathcal{N} ||_{ \sigma } \leq \mathcal{N} \otimes_1 \bar{\mathbf{x}}_1 \cdots \otimes_N \bar{\mathbf{x}}_N + \underbrace{ \left( \epsilon N + \epsilon^2 \binom{N}{2} + \cdots + \epsilon^N \binom{N}{N} \right) }_{ (\star) } || \mathcal{N} ||_{ \sigma } .
\end{equation*}
Furthermore, we choose $\epsilon = \frac{ \log \frac{3}{2} }{ N }$ and estimate the above $(\star)$ term as follows
\begin{equation*}
  (\star) \leq \epsilon N + \frac{ (\epsilon N)^2 }{ 2! } + \cdots + \frac{ (\epsilon N )^N }{ N! } \leq \mathrm{e}^{ \epsilon N } - 1 = \frac{1}{2}.
\end{equation*}
Hence
\begin{equation*}
  || \mathcal{N} ||_{ \sigma } \leq 2 \max\limits_{ \bar{\mathbf{x}}_k \in \mathcal{C}_k, k = 1, \cdots, N } \mathcal{N} \otimes_1 \bar{\mathbf{x}}_1 \cdots \otimes_N \bar{\mathbf{x}}_N.
\end{equation*}

Using the property of $\epsilon$-covering and Eq.~\ref{eq:auxiliary_bound} we can derive the following inequality for any $\lambda \geq 0$
\begin{align*}
  \Pr ( || \mathcal{N} ||_{\sigma} \geq \lambda ) & \leq \Pr ( 2 \max\limits_{ \bar{\mathbf{x}}_k \in \mathcal{C}_k, k = 1, \cdots, N } \mathcal{N} \otimes_1 \bar{\mathbf{x}}_1 \cdots \otimes_N \bar{\mathbf{x}}_N \geq \lambda ) \\
  & \leq \sum\limits_{ \bar{\mathbf{x}}_k \in \mathcal{C}_k, k = 1, \cdots, N } \leq \left( \frac{2}{\epsilon} \right)^{ \sum\limits_{k=1}^N d_k } 2 \mathrm{e}^{ - \frac{ p \lambda^2 }{ 8 } }.
\end{align*}

Setting $ \Pr ( || \mathcal{N} ||_{\sigma} \geq \lambda ) = \delta $, the spectral norm of the noise tensor $\mathcal{N}$ can be bounded by
\begin{equation}
  || \mathcal{N} ||_{ \sigma } \leq \sqrt{ \frac{8}{p} \left( \log ( \frac{ 2N }{ N_0 } ) \sum\limits_{k=1}^N d_k  + \log \frac{2}{\delta} \right) }, \quad N_0 := \log \frac{3}{2}
\end{equation}
with probability at least $1-\delta$ if the sample probability satisfies $p \geq 0.22$.
\end{proof}

Using $ || \mathcal{N}_r ||_{ \sigma } = || \mathcal{N} ||_{ \sigma }$, and $ || \mathcal{N}_r ||_F \leq \sqrt{r}  || \mathcal{N}_r ||_{ \sigma } $ we can estimate the norms of the truncated tensor SVD of the noise tensor.
\begin{appendix_lemma}
  \begin{align*}
    || \mathcal{N}_r ||_{ \sigma } & \leq \sqrt{ \frac{8}{p} \left( \log ( \frac{ 2N }{ N_0 } ) \sum\limits_{k=1}^N d_k  +\log \frac{2}{\delta} \right)  } \\
    || \mathcal{N}_r ||_F & \leq  \sqrt{ r \frac{8}{p} \left( \log ( \frac{ 2N }{ N_0 } ) \sum\limits_{k=1}^N d_k  +\log \frac{2}{\delta} \right) },
  \end{align*}
where $ N_0 = \log \frac{3}{2}$ and the sample probability should satisfy $p \geq 0.22$.
\label{lemma:norm_low_rank_noise}
\end{appendix_lemma}

Now we are able to determine the sample probability, such that the error ratio $ \frac{ || \mathcal{A} - \hat{ \mathcal{A}}_r ||_F }{ || \mathcal{A} ||_F } $ is bounded.
\begin{appendix_theorem}[Theorem 1 in the main text]
  Let $ \mathcal{A} \in \{ 0, 1 \}^{ d_1 \times d_2 \times \cdots \times d_N}$. Suppose that $\mathcal{A}$ can be well approximated by its $r$-rank tensor SVD $\mathcal{A}_r$. Using the subsampling scheme defined in Eq.~\ref{eq:subsample_method} with the sample probability $ p \geq \max \{ 0.22,  8 r \left( \log ( \frac{ 2N }{ N_0 } ) \sum\limits_{k=1}^N d_k  +\log \frac{2}{\delta} \right) /  (  \tilde{\epsilon} || \mathcal{A} ||_F )^2 \} $, $ N_0 = \log \frac{3}{2}$, then the original tensor $\mathcal{A}$ can be reconstructed from the truncated tensor SVD of the subsampled tensor $ \hat{ \mathcal{A} } $. The error satisfies $ || \mathcal{A} - \hat{ \mathcal{A}}_r ||_F  \leq  \epsilon || \mathcal{A} ||_F $ with probability at least $1 - \delta$, where $\epsilon$ is a function of $\tilde{ \epsilon }$. Especially, $\tilde{ \epsilon }$ together with the sample probability controls the norm of the noise tensor.
  \label{theorem:bound_by_rank}
\end{appendix_theorem}

\begin{proof}
  Suppose tensor $\mathcal{A}$ can be well approximated by its $r$-rank tensor SVD, in a sense that $ || \mathcal{A} - \mathcal{A}_r || \leq \epsilon_0 || \mathcal{A} ||_F $ for some small $\epsilon_0 > 0$. According to Lemma A ~\ref{lemma:norm_low_rank_noise} if we want the Frobenius norm of the noise tensor $\mathcal{N}_r$ to be bounded by $ \tilde{\epsilon} || \mathcal{A} ||_F $ with $\tilde{\epsilon} > 0$, then the sample probability should satisfy $ p \geq \{ 0.22,  \frac{ 8 r \left( \log ( \frac{ 2N }{ N_0 } ) \sum\limits_{k=1}^N d_k  +\log \frac{2}{\delta} \right) }{ (  \tilde{\epsilon} || \mathcal{A} ||_F )^2 } \}$.

Using Eq.~\ref{eq:reconstruction_error} we have
  \begin{equation*}
    || \mathcal{A} - \hat{ \mathcal{A} }_r ||_F \leq 2 \epsilon_0 || \mathcal{A} ||_F + 2 \sqrt{\epsilon_0} || \mathcal{A} ||_F + 2 \sqrt{ \tilde{ \epsilon} } || \mathcal{A} ||_F + \tilde{\epsilon} || \mathcal{A} ||_F = \epsilon || \mathcal{A} ||_F,
  \end{equation*}
where $ \epsilon := 2 ( \epsilon_0 + \sqrt{ \epsilon_0 } +  \sqrt{ \tilde{ \epsilon } } )  + \tilde{ \epsilon }$.
\end{proof}

Note that in the case where $\mathcal{A}$ is a two-dimensional matrix, the sample probability derived in~\cite{achlioptas2007fast} reads $ \mathcal{O} ( \frac{d_1 + d_2}{ || \mathcal{A} ||_F^2} )$. This corresponds the high-dimensional tensor case.

For the later use in the quantum algorithm, instead of considering low-rank approximation of the subsampled tensor, we study the tensor SVD with projected singular values, denoted as $ \hat{\mathcal{A}}_{ |\cdot| \geq \tau }$. This notation denotes that subsampled tensor $\hat{\mathcal{A}}$ is projected onto the eigenspaces with absolute singular values larger than a threshold. Later, it will be also referred to as the projected tensor SVD of $ \hat{ \mathcal{A} } $ with threshold $\tau$. The following theorem discusses the choice of sample probability and threshold $\tau$, such that the error ratio $ \frac{ || \mathcal{A} - \hat{ \mathcal{A} }_{ |\cdot| \geq \tau} ||_F }{ || \mathcal{A} ||_F }$ is bounded.

\begin{appendix_theorem}[Theorem 2 in the main text]
  Let $ \mathcal{A} \in \{ 0, 1 \}^{ d_1 \times d_2 \times \cdots \times d_N}$. Suppose that $\mathcal{A}$ can be well approximated by its $r$-rank tensor SVD $\mathcal{A}_r$. Using the subsampling scheme defined in Eq.~\ref{eq:subsample_method} with the sample probability $ p \geq \max \{ 0.22, p_1 := \frac{ l_1 C_0 }{ ( \tilde{\epsilon} || \mathcal{A} ||_F )^2 }, p_2 := \frac{ r C_0 }{ ( \tilde{\epsilon} || \mathcal{A} ||_F )^2 }, p_3 := \frac{ \sqrt{2 r C_0 } }{ \epsilon_1 \tilde{\epsilon} || \mathcal{A} ||_F } \}$, with $\quad C_0 =  8 \left( \log ( \frac{ 2N }{ N_0 } ) \sum_{k=1}^N d_k  +\log \frac{2}{\delta} \right) $, $ N_0 =\log \frac{3}{2}$, where $l_1$ denotes the largest index of singular values of tensor $ \hat{\mathcal{A}} $ with $ \sigma_{l_1} \geq \tau $, and choosing the threshold as $ 0 < \tau \leq \frac{ \sqrt{2 C_0 }}{ p \tilde{\epsilon} }$, then the original tensor $\mathcal{A}$ can be reconstructed from the projected tensor SVD of $ \hat{ \mathcal{A} } $. The error satisfies $ || \mathcal{A} - \hat{ \mathcal{A}}_{ |\cdot| \geq \tau } ||_F  \leq  \epsilon || \mathcal{A} ||_F $ with probability at least $1 - \delta$, where $\epsilon$ is a function of $\tilde{ \epsilon }$ and $\epsilon_1$. Especially, $\tilde{ \epsilon }$ together with $p_1$ and $p_2$ determine the norm of noise tensor and $ \epsilon_1 $ together with $p_3$ control the value of $ \hat{ \mathcal{A} } $'s singular values that are located outside the projection boundary.
  \label{theorem:bound_by_threshold}
\end{appendix_theorem}

\begin{proof}
  Suppose tensor $\mathcal{A}$ can be well approximated by its $r$-rank tensor SVD, in a sense that $ || \mathcal{A} - \mathcal{A}_r || \leq \epsilon_0 || \mathcal{A} ||_F $ for some small $\epsilon_0 > 0$. Define the threshold as $ \tau := \kappa  || \hat{ \mathcal{A} } ||_F > 0 $ for some $ \kappa > 0$. Let $l_1$ denote the largest index of singular values of tensor $ \hat{\mathcal{A}} $ with $ \sigma_{l_1} \geq \kappa || \hat{ \mathcal{A} }  ||_F $, and let $l_2$ denote the smallest index of singular values of tensor $ \hat{\mathcal{A}} $ with $ \sigma_{ l_2 } \leq - \kappa || \hat{ \mathcal{A} } ||_F $. If the threshold $\tau$ is large enough, we only need to consider the case $ l_1 \ll l_2$. Moreover, we have the following constrain for $l_1$ and $\kappa$:
  \begin{equation}
    l_1 \cdot \sigma_{l_1}^2 \leq || \hat{ \mathcal{A} }_{l_1} ||_F^2 \leq || \hat{ \mathcal{A} } ||_F^2 \quad \Rightarrow \quad l_1 \cdot \kappa^2 \leq 1.
    \label{eq:kl_constraint}
  \end{equation}

Suppose that the tensor $\hat{ \mathcal{A} }$ can be well approximated by the tensor SVD with rank $R$ which is written as $ \hat{ \mathcal{A}_R } $. Note that the rank $R$ can be much larger than $r$. We first bound $ || \mathcal{A} - \hat{\mathcal{A}}_{ |\cdot| \geq \tau } ||_F $ as follows
  \begin{align*}
    || \mathcal{A} - \hat{\mathcal{A}}_{ |\cdot| \geq \tau } ||_F & \approx || \mathcal{A} - \hat{\mathcal{A}}_{ [0, l_1] \cup [l_2, R]} ||_F = || \mathcal{A} - ( \hat{\mathcal{A}}_R - \hat{\mathcal{A}}_{l_2} + \hat{\mathcal{A}}_{l_1} ) ||_F \\
    & \leq || \mathcal{A} - \hat{\mathcal{A}}_{l_1} ||_F + || \hat{\mathcal{A}}_{l_2} - \hat{\mathcal{A}}_R ||_F = || \mathcal{A} - \hat{\mathcal{A}}_{l_1} ||_F + || \mathcal{A} - \mathcal{A} + \hat{\mathcal{A}}_{l_2} - \hat{\mathcal{A}}_R ||_F \\
    & \leq  || \mathcal{A} - \hat{\mathcal{A}}_{l_1} ||_F  + || \mathcal{A} - \hat{\mathcal{A}}_{R} ||_F + || \mathcal{A} - \hat{\mathcal{A}}_{l_2} ||_F \\
    & \leq 3 || \mathcal{A} - \hat{\mathcal{A}}_{l_1} ||_F.
  \end{align*}

Assume $l_1 \ll l_2$ and we only distinguish two cases: $ l_2 \gg l_1 \geq r$ and $l_1 < r \ll l_2$.

Suppose $l_1 \geq r$, we have
  \begin{align*}
    || \mathcal{A} - \hat{ \mathcal{A} }_{ |\cdot| \geq \tau} ||_F & \leq 3 || \mathcal{A} - \hat{ \mathcal{A} }_{l_1} ||_F  \\  & \stackrel{(1)}{ \leq } 3 ( 2 || \mathcal{A} - \mathcal{A}_{l_1} ||_F + 2 \sqrt{ || \mathcal{A}_{l_1} ||_F || \mathcal{A} - \mathcal{A}_{l_1} ||_F } + 2 \sqrt{ || \mathcal{N}_{l_1} ||_F || \mathcal{A}_{l_1} ||_F } + || \mathcal{N}_{l_1} ||_F )   \\
    & \stackrel{(2)}{ \leq }  3 ( 2 || \mathcal{A} - \mathcal{A}_r ||_F + 2 \sqrt{ || \mathcal{A} ||_F || \mathcal{A} - \mathcal{A}_{l_1} ||_F } + 2 \sqrt{ || \mathcal{N}_{l_1} ||_F || \mathcal{A} ||_F } + || \mathcal{N}_{l_1} ||_F ),
  \end{align*}
where inequality $(1)$ is given by Eq.~\ref{eq:reconstruction_error} and $(2)$ uses $ || \mathcal{A}_{l_1} ||_F  \leq || \mathcal{A} ||_F $.

According to Lemma A ~\ref{lemma:norm_low_rank_noise} if we want the Frobenius norm $ || \mathcal{N}_{l_1} ||_F $ to be bounded by $ \tilde{\epsilon} || \mathcal{A} ||_F $ with $\tilde{\epsilon} > 0$, then the sample probability should satisfy $ p \geq \max \{ 0.22, p_1 := \frac{l_1  \ C_0 }{ ( \tilde{\epsilon} || \mathcal{A} ||_F )^2 } \} $ where the constant is defined as $ C_0 :=  8 \left( \log ( \frac{ 2N }{ N_0 } ) \sum\limits_{k=1}^N d_k  +\log \frac{2}{\delta} \right) $ (see Lemma A~\ref{lemma:norm_low_rank_noise}). Finally, under this sample condition we have $ || \mathcal{A} - \hat{ \mathcal{A} }_{ |\cdot| \geq \tau } ||_F \leq 3 ( 2 \epsilon_0 + 2 \sqrt{\epsilon_0} + 2 \sqrt{ \tilde{\epsilon} } + \tilde{\epsilon}) || \mathcal{A} ||_F $ for $l_1 \geq r$.

Before considering the case $l_1 < r \ll l_2 $ we first estimate the Frobenius norm of subsampled tensor. $ || \hat{ \mathcal{A} } ||_F^2 $ can be written as a sum of random variables $X_l :=  \hat{\mathcal{A}}_l^2 $ for $l = 1, \cdots, d_1 d_2 \cdots d_N$ using a bijection of indices, namely $ X := || \hat{ \mathcal{A} } ||_F^2 = \sum_l X_l $. Moreover, $ \mathbb{E} [ X_l ] = \frac{1}{p} \mathcal{A}_l^2 $ and $ \mathbb{E} [X] = \frac{1}{p} || \mathcal{A} ||_F^2 $. According to the Chernoff bound
\begin{equation}
  \Pr ( | X - \mathbb{E} [X] | \geq \delta \mathbb{E} [X] ) \leq 2 \mathrm{e}^{ - \frac{\mathbb{E} [X] \delta^2 }{ 3 } }  \quad \text{for all } \quad 0 < \delta < 1,
\end{equation}
we have $ \Pr ( || \hat{\mathcal{A} } ||_F^2 \geq \frac{ 1 + \delta }{p} || \mathcal{A} ||_F^2 ) \leq 2 \mathrm{e}^{ - \frac{ || \mathcal{A} ||_F^2 \delta^2 }{3p} } $ for $ \delta \in (0, 1)$. Hence $ || \hat{\mathcal{A}} ||_F \leq \sqrt{ \frac{2}{p} } || \mathcal{A} ||_F$ is satisfied with high probability.

In the following, we study the case $l_1 < r \ll l_2 $ and fix the sample probability $p$ temporarily. It gives
\begin{align}
    & || \mathcal{A} - \hat{ \mathcal{A} }_{ |\cdot| \geq \tau} ||_F  \leq 3 || \mathcal{A} - \hat{ \mathcal{A} }_{l_1} ||_F  \leq 3 ( || \mathcal{A} - \hat{ \mathcal{A} }_r ||_F + || \hat{ \mathcal{A} }_r - \hat{ \mathcal{A} }_{l_1} ||_F  )  \nonumber \\
    & \ \stackrel{(1)}{ \leq } 3 ( 2 || \mathcal{A} - \mathcal{A}_r ||_F + 2 \sqrt{ || \mathcal{A}_r ||_F || \mathcal{A} - \mathcal{A}_r ||_F }  + 2 \sqrt{ || \mathcal{N}_r ||_F || \mathcal{A}_r ||_F } + || \mathcal{N}_r ||_F + || \hat{ \mathcal{A} }_r - \hat{ \mathcal{A} }_{l_1} ||_F  ) \nonumber  \\
    & \ \stackrel{(2)}{ \leq } 3 ( 2 || \mathcal{A} - \mathcal{A}_r ||_F + 2 \sqrt{ || \mathcal{A} ||_F || \mathcal{A} - \mathcal{A}_r ||_F }  + 2 \sqrt{ || \mathcal{N}_r ||_F || \mathcal{A} ||_F } + || \mathcal{N}_r ||_F + \underbrace{ \sqrt{ \frac{2r}{p} } \kappa || \mathcal{A} ||_F }_{ ( \star ) }  ),
    \label{eq:star_term}
\end{align}
where inequality $(1)$ is given by Eq.~\ref{eq:reconstruction_error} and $(2)$ uses the following estimation
\begin{equation*}
  || \hat{ \mathcal{A} }_r - \hat{ \mathcal{A} }_{l_1} ||_F \leq \sqrt{r - l_1} \tau \leq \sqrt{r} \tau = \sqrt{r} \kappa || \hat{ \mathcal{A} } ||_F \leq \sqrt{ \frac{ 2r }{p}  } \kappa || \mathcal{A} ||_F.
\end{equation*}
Similarly, if we want the Frobenius norm $ || \mathcal{N}_r ||_F $ to be bounded by $ \tilde{\epsilon} || \mathcal{A} ||_F $ with $\tilde{\epsilon} > 0$, then the sample probability should satisfy $ p \geq \max \{ 0.22, p_1 := \frac{r  \ C_0 }{ ( \tilde{\epsilon} || \mathcal{A} ||_F )^2 } \} $ according to Lemma A ~\ref{lemma:norm_low_rank_noise}. In order to choose $\kappa$, we fix the sample probability $p$ temporarily and use the constraint Eq.~\ref{eq:kl_constraint}. It gives
\begin{equation}
    l_1 < r = \frac{ p ( \tilde{\epsilon} || \mathcal{A} ||_F )^2 }{ C_0 } \quad \Rightarrow \quad \kappa^2 \leq \frac{ C_0 }{ p ( \tilde{\epsilon} || \mathcal{A} ||_F )^2 }.
    \label{eq:kappa_ineq}
\end{equation}

We can further control the sum of singular values that are located outside the projection boundary by requiring $(\star) \leq \epsilon_1 || \mathcal{A} ||_F $ for some small $\epsilon_1 > 0$ in Eq.~\ref{eq:star_term}. Plug the above inequality of $\kappa$ into the $( \star )$ term we obtain another condition for the sample probability
\begin{equation}
  \sqrt{ \frac{2r}{p} } \kappa \leq \epsilon_1 \quad \Rightarrow \quad p \geq \frac{ \sqrt{ 2 r C_0 } }{ \epsilon_1 \tilde{ \epsilon } || \mathcal{A} ||_F } := p_3.
\end{equation}
Therefore, in the case $ l_1 < r \ll l_2 $ if $ p \geq \max \{ 0.22, p_2 = \frac{ r C_0 }{ ( \tilde{\epsilon} || \mathcal{A} ||_F )^2 }, p_3 = \frac{ \sqrt{2 r C_0 } }{ \epsilon_1 \tilde{\epsilon} || \mathcal{A} ||_F } \}$ we have $ || \mathcal{A} - \hat{ \mathcal{A} }_{ |\cdot| \geq \tau} ||_F  \leq \epsilon || \mathcal{A} ||_F $, where $ \epsilon := 3 (2 \epsilon_0 + 2 \sqrt{ \epsilon_0 } + 2 \sqrt{ \tilde{ \epsilon } } + \tilde{\epsilon} + \epsilon_1 ) $.

In summary, combine two situations we have $ || \mathcal{A} - \hat{ \mathcal{A} }_{ |\cdot| \geq \tau } ||_F \leq  \epsilon || \mathcal{A} ||_F $, where $ \epsilon := 3 ( 2 \epsilon_0 + 2 \sqrt{ \epsilon_0 } + 2 \sqrt{ \tilde{\epsilon} } + \tilde{\epsilon} + \epsilon_1 ) $ if the sample probability is chosen as
\begin{equation*}
  p \geq \max \{ 0.22, p_1 = \frac{ l_1 C_0 }{ ( \tilde{\epsilon} || \mathcal{A} ||_F )^2 },  p_2 = \frac{ r C_0 }{ ( \tilde{\epsilon} || \mathcal{A} ||_F )^2 }, p_3 = \frac{ \sqrt{2 r C_0 } }{ \epsilon_1 \tilde{\epsilon} || \mathcal{A} ||_F } \}.
\end{equation*}
Moreover, the threshold can be determined from the following approximation after choosing the sample probability:
\begin{equation*}
  \tau = \kappa || \hat{\mathcal{A}} ||_F \leq \sqrt{ \frac{ C_0 }{ p \tilde{\epsilon}^2 } } \frac{ || \hat{\mathcal{A}} ||_F }{ || \mathcal{A} ||_F }   \leq \frac{ \sqrt{2 C_0} }{ p \tilde{\epsilon} },
\end{equation*}
where the inequality is derived by using Eq.~\ref{eq:kappa_ineq} and $ || \hat{\mathcal{A}} ||_F \leq \sqrt{ \frac{2}{p} } || \mathcal{A} ||_F$.

\end{proof}
The above estimation on the error bound in the case of projected tensor SVD is crucial for the quantum algorithm since quantum singular value projection depends only on the positive threshold defined for the singular values.

\subsection{Data Structure}

\begin{appendix_theorem}
  \cite{prakash2014quantum} Let $ \mathbf{x} \in \mathbb{R}^R $ be a real-valued vector. The quantum state $ \ket{x} = \frac{1}{ || \mathbf{x} ||_2 } \sum\limits_{i=1}^R x_i \ket{i}$ can be prepared using $ \lceil \log_2 R \rceil $ qubits in time $ \mathcal{O} ( \log_2 R ) $.
  \label{theorem:data_structure_appendix}
\end{appendix_theorem}

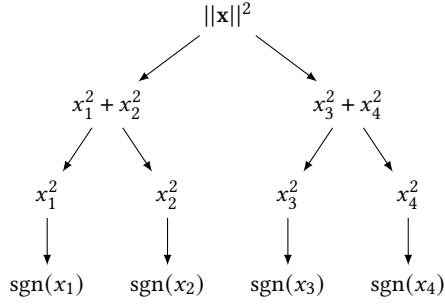
\begin{figure}[htp]
\centering
\begin{tikzpicture}[scale=.8, edge from parent/.style={draw,-latex},
  level 1/.style={sibling distance=4.cm},
  level 2/.style={sibling distance=2.cm},
  every  node/.style = {scale=1.}]
  \node { $ || \mathbf{x} ||^2 $ }
    child {node { $ x_1^2 + x_2^2 $ }
      child {node { $ x_1^2 $ }
        child { node { $ \mathrm{sgn} (x_1) $ } } }
      child {node { $ x_2^2 $ }
        child { node { $ \mathrm{sgn} (x_2) $ } } }
    }
    child {node { $ x_3^2 + x_4^2 $ }
      child {node { $ x_3^2 $ }
        child { node { $ \mathrm{sgn} (x_3) $ } } }
      child {node { $ x_4^2 $ }
        child { node { $ \mathrm{sgn} (x_4) $ } } }
     };
\end{tikzpicture}
  \caption{Classical memory structure with quantum access for creating the quantum state $ \ket{x} = x_1 \ket{00} + x_2 \ket{01} + x_3 \ket{10} + x_4 \ket{11}$. }
  \label{fig:tree_data_structure}
\end{figure}

Theorem A ~\ref{theorem:data_structure_appendix} claims that there exist a classical memory structure and a quantum algorithm which can load classical data into a quantum state with acceleration. Figure~\ref{fig:tree_data_structure} illustrates a simple example. Given an $R=4$ dimensional real-valued vector, the quantum state $ \ket{x} = x_1 \ket{00} + x_2 \ket{01} + x_3 \ket{10} + x_4 \ket{11}$ can be prepared by querying the classical memory structure and applying $3$ controlled rotations.

Let us assume that $ \mathbf{x} $ is normalized, namely $ || \mathbf{x} ||_2 = 1 $. The quantum state $\ket{x}$ is created from the initial state $ \ket{0} \ket{0}$ by querying the memory structure from the root to the leaf. The first rotation is applied on the first qubit, giving
\begin{equation*}
  ( \cos \theta_1 \ket{0} + \sin \theta_1 \ket{1} ) \ket{0} = ( \sqrt{x_1^2 + x_2^2 } \ket{0} + \sqrt{x_3^2 + x_4^2} \ket{1} ) \ket{0},
\end{equation*}
where $ \theta_1 := \tan^{-1} \sqrt{ \frac{ x_3^2 + x_4^2 }{ x_1^2 + x_2^2 } } $. The second rotation is applied on the second qubit conditioned on the state of qubit $1$. It gives
\begin{equation*}
  \sqrt{x_1^2 + x_2^2 } \ket{0} \frac{1}{ \sqrt{x_1^2 + x_2^2 } } ( |x_1| \ket{0} + |x_2| \ket{1} ) + \sqrt{x_3^2 + x_4^2 } \ket{1} \frac{1}{ \sqrt{x_3^2 + x_4^2 } } ( |x_3| \ket{0} + |x_4| \ket{1} ).
\end{equation*}
The last rotation loads the signs of coefficients conditioned on qubits $1$ and $2$. In general, an $R$-dimensional real-valued vector needs to be stored in a classical memory structure with $ \lceil \log _2 R  \rceil + 1$ layers. The data vector can be loaded into a quantum state using $ \mathcal{O} ( \log _2 R ) $ non-trivial controlled rotations.

The above simple example of quantum Random Access Memory for generating quantum state from a real-valued vector can be generalized to quantum access of other more complicated data structures, e.g., matrices, tensors.

\subsection{Simulation of the unitary operator $ \mathrm{e}^{ - i t \tilde{\rho}_{ \hat{\chi}^{\dagger} \hat{ \chi } } } $}

Before proving Lemma A~\ref{lemma:simulation_operator} of unitary operator simulation in the main text, we give the following auxiliary Lemma. The difficulty of simulating a unitary operator $\mathrm{e}^{ - i \rho t}$ up to time $t$ is to efficiently exponentiate the density matrix $\rho$. In~\cite{lloyd1996universal}, Lloyd suggested an efficient algorithm for Hamiltonian simulation using a tensor product structure. In particular, the unitary operator $\mathrm{e}^{ - i \rho \Delta t }$ with short simulation time $\Delta t$ can be constructed via a simple swap operator.

\begin{appendix_lemma}
  Let $\rho$ and $\sigma$ be density matrices, and $S$ a swap operator such that $S \ket{x} \ket{y} = \ket{y} \ket{x} $. Then for an infinitesimal simulation step $\Delta t$ we have $ \mathrm{e}^{ - i \rho \Delta t } \sigma \mathrm{e}^{ i \rho \Delta t } =  \mathrm{tr}_1 \{ \mathrm{e}^{ - i S \Delta t } \rho \otimes \sigma \mathrm{e}^{ i S \Delta t } \} $ up to the first order of $\Delta t$, where $\mathrm{tr}_1$ is a partial trace applied on the first subsystem of the tensor product structure.
  \label{lemma:small_step_hamiltonian_simulation}
\end{appendix_lemma}

\begin{proof}
First note that density matrices $\rho$ and $\sigma$ can be written in the eigenbasis as $ \rho = \sum\limits_i \ket{ \rho_i } \bra{ \rho_i } $ and $ \sigma = \sum\limits_j \ket{ \sigma_j } \bra{ \sigma_j } $. Moreover, for $ \Delta t \rightarrow 0 $ we have approximations $ \mathrm{e}^{ - i S \Delta t } \approx \cos \Delta t \ I - i \sin \Delta t \ S $ and $ \mathrm{e}^{ i S \Delta t } \approx \cos \Delta t \ I + i \sin \Delta t \ S $, where $I$ denotes the identity operator.

Hence
\begin{align*}
  \mathrm{tr}_1 \{ \mathrm{e}^{ - i S \Delta t } \rho \otimes \sigma \mathrm{e}^{ i S \Delta t }  \} = & \mathrm{tr}_1 \{ ( \cos \Delta t \ I - i \sin \Delta t \ S ) ( \sum\limits_{ij} \ket{\rho_i} \ket{\sigma_j} \bra{\rho_i} \bra{\sigma_j} ) ( \cos \Delta t \ I + i \sin \Delta t \ S )  \}  \\
  = & \mathrm{tr}_1 \{ \sum\limits_{ij} [  \cos^2 \Delta t \ket{ \rho_i } \ket{ \sigma_j } \bra{ \rho_i } \bra{ \sigma_j } - i \sin \Delta t \cos \Delta t S \ket{ \rho_i } \ket{ \sigma_j } \bra{ \rho_i } \bra{ \sigma_j }  \\
  & \quad \quad + i \cos \Delta t \sin \Delta t \ket{ \rho_i } \ket{ \sigma_j } \bra{ \rho_i } \bra{ \sigma_j } S^{\dagger}  + \sin^2 \Delta t S \ket{ \rho_i } \ket{ \sigma_j } \bra{ \rho_i } \bra{ \sigma_j } S^{\dagger}  ] \}
\end{align*}

Recall that $S \ket{ \rho_i } \ket{ \sigma_j } = \ket{ \sigma_j } \ket{ \rho_i } $ and $ \bra{ \rho_i } \bra{ \sigma_j } S^{ \dagger } = \bra{ \sigma_j } \bra{ \rho_i } $. Applying the swap operator $S$ gives
\begin{equation*}
  \mathrm{tr}_1 \{ \mathrm{e}^{ - i S \Delta t } \rho \otimes \sigma \mathrm{e}^{ i S \Delta t }  \} \approx \sigma - i \Delta t [ \sum\limits_{ij}  \braket{ \rho_i | \sigma_j  } \ket{ \rho_i } \bra{ \sigma_j } - \sum\limits_{ij}   \braket{ \sigma_j | \rho_i } \ket{ \sigma_j } \bra{ \rho_i } ] + \mathcal{O} ( \Delta t ^2 ),
\end{equation*}
where we used $\cos \Delta t \approx 1$ and $ \sin \Delta t \approx \Delta t$ as $ \Delta t \rightarrow 0$. The commutator of two operators is defined as
\begin{equation*}
  [ \rho, \sigma ] :=  \rho \sigma - \sigma \rho = \sum\limits_{ij} \braket{ \rho_i | \sigma_j } \ket{ \rho_i } \bra{ \sigma_j } - \sum\limits_{ij}  \braket{ \sigma_j | \rho_i } \ket{ \sigma_j } \bra{ \rho_i }
\end{equation*}
and we finally have
\begin{equation}
  \mathrm{tr}_1 \{ \mathrm{e}^{ - i S \Delta t } \rho \otimes \sigma \mathrm{e}^{ i S \Delta t }  \} = \sigma - i \Delta t [ \rho, \sigma ] + \mathcal{ O } ( \Delta t^2 ).
  \label{eq:trace_swap}
\end{equation}

On the other hand, applying the limits $ \lim\limits_{ \Delta t \rightarrow 0} \mathrm{e}^{ - i \rho \Delta t } = I - i \rho \Delta t $ and $ \lim\limits_{ \Delta t \rightarrow 0} \mathrm{e}^{ i \rho \Delta t } = I + i \rho \Delta t $ we can derive
\begin{align*}
  \mathrm{e}^{ - i \rho \Delta t } \sigma \mathrm{e}^{ i \rho \Delta t } & \approx ( I - i \Delta t \sum\limits_i \ket{\rho_i} \bra{\rho_i} ) \sum\limits_j \ket{\sigma_j} \bra{\sigma_j} ( I + i \Delta t \sum\limits_i \ket{\rho_i} \bra{\rho_i} ) \\
  = & \sigma - i \Delta t [ \rho, \sigma ] + \mathcal{O} (\Delta t^2).
\end{align*}

In summary, we have $ \mathrm{e}^{ - i \rho \Delta t } \sigma \mathrm{e}^{ i \rho \Delta t } =  \mathrm{tr}_1 \{ \mathrm{e}^{ - i S \Delta t } \rho \otimes \sigma \mathrm{e}^{ i S \Delta t } \} $ up to th first order of $\Delta t$. The above proof indicates that we can construct the unitary operator $ \mathrm{e}^{ - i \rho t} $ and act on the density $\sigma$ by repeatedly applying simple operations $ \mathrm{e}^{ - i S \Delta t } \approx I - i S \Delta t $ on the tensor product state $ \rho \otimes \sigma $ in $ n = \frac{t}{ \Delta t } $ steps.
\end{proof}

\begin{appendix_lemma}[Lemma 3 in the main text]
  Unitary operator $ \mathrm{e}^{ - i t \tilde{\rho}_{ \hat{\chi}^{\dagger} \hat{ \chi } } }$ can be applied to any quantum state, where $ \tilde{\rho}_{ \hat{\chi}^{\dagger} \hat{ \chi } } := \frac{ \rho_{ \hat{\chi}^{\dagger} \hat{ \chi } } }{ d_2 d_3 } $, up to simulation time $t$. The total number of steps for simulation is $ \mathcal{O} ( \frac{ t^2 }{ \epsilon } T_{\rho} )$, where $\epsilon$ is the desired accuracy, and $T_{\rho}$ is the time for accessing the density matrix.
  \label{lemma:simulation_operator}
\end{appendix_lemma}
\begin{proof}
  The proof uses the dense matrix exponentiation method proposed in~\cite{rebentrost2018quantum} which was developed from~\cite{lloyd1996universal}. Recall that $ \rho_{ \hat{\chi}^{\dagger} \hat{ \chi } } = \sum\limits_{i_2 i_3 i_2' i_3'} \mathcal{C}_{ i_2 i_3 i_2' i_3' } \ket{i_2 i_3} \bra{i_2' i_3'}$, where $ \mathcal{C}_{ i_2 i_3 i_2' i_3' } = \sum\limits_{i_1} \hat{ \chi }^{ \dagger}_{ i_1, i_2 i_3 } \hat{ \chi }_{ i_1, i_2' i_3'}$. For the sake of simplicity, we rewrite $ \rho_{ \hat{\chi}^{\dagger} \hat{ \chi } }  $ as $A \in \mathbb{R}^{ N^2 \times N^2 }$, where $N := d_2 d_3$. Suppose that the unitary operator needs to be applied on the quantum state $\ket{x}$ whose density matrix reads $\sigma := \ket{x} \bra{x}$. Then follow the method in~\cite{rebentrost2018quantum}, we first create a modified swap operator
  \begin{equation*}
    S_A = \sum\limits_{j, k = 1}^N A_{jk} \ket{k} \bra{j} \otimes \ket{j} \bra{k},
  \end{equation*}
and another auxiliary density matrix $ \mu = \ket{ \vec{1} } \bra{ \vec{1} }$, with $ \ket{ \vec{1} } := \frac{1}{\sqrt{N}} \sum\limits_{k=1}^N \ket{k} $. Consider the evolution of the system $ \mu \otimes \sigma  $ under the unitary operator $ \mathrm{e}^{ - i S_A \Delta t}$ for a small step $\Delta t$. With Lemma A~\ref{lemma:small_step_hamiltonian_simulation} it can be shown that
  \begin{equation*}
    \mathrm{tr}_1 \{  \mathrm{e}^{ - i S_A \Delta t } \mu \otimes \sigma \mathrm{e}^{ i S_A \Delta t} \} \approx \mathrm{e}^{ - i \frac{A}{N} \Delta t} \sigma \mathrm{e}^{ i \frac{A}{N} \Delta t}.
  \end{equation*}
Moreover, repeated applications of $\mathrm{e}^{ - i S_A \Delta t }$, say $n$ times with $ t := n \Delta t$, on the bigger system $\mu \otimes \sigma$ can give $ \mathrm{e}^{ - i \frac{A}{N} t} \sigma \mathrm{e}^{ i \frac{A}{N} t} $ with is the density matrix of the quantum state $ \mathrm{e}^{ - i \frac{A}{N} t} \ket{x} $. In other words, we can simulate the unitary operator $ \mathrm{e}^{ - i t \tilde{\rho}_{ \hat{\chi}^{\dagger} \hat{ \chi } } }  $ with $ \tilde{\rho}_{ \hat{\chi}^{\dagger} \hat{ \chi } } := \frac{ \rho_{ \hat{\chi}^{\dagger} \hat{ \chi } } }{ d_2 d_3 } $.

Furthermore, \cite{rebentrost2018quantum} shows that given $t$ and the required accuracy $\epsilon$, the step size $\Delta t$ should be small enough, such that $ n = \mathcal{O} (\frac{t^2}{ \epsilon})$. In addition, the quantum access for obtaining the density $ \rho_{ \hat{\chi}^{\dagger} \hat{\chi} } $ and creating the modified swap operator requires $ T_{\rho} = \mathcal{O} ( \mathrm{polylog} (d_1 d_2 d_3) ) $ steps. In summary, the total run time for simulating $ \mathrm{e}^{ - i t \tilde{\rho}_{ \hat{\chi}^{\dagger} \hat{ \chi } } } \ket{x}  $ is $ nT_{\rho} = \mathcal{O} ( \frac{t^2}{\epsilon} \mathrm{polylog} (d_1 d_2 d_3) )$.
\end{proof}

\newpage
\bibliographystyle{plainnat}
\bibliography{acm_qsvd}

\end{document}